\documentclass[uxseAMS, usenatbib]{mn2e}
\usepackage{graphicx}     
\usepackage{amsmath,environ}
\usepackage{color}

\title[Star/galaxy separation at faint magnitudes, applied to DES.]
{Star/galaxy separation at faint magnitudes: \\Application to a simulated Dark Energy Survey}
\author[M. T. Soumagnac et al.]
{M. T. Soumagnac $^{1,2}$\thanks{E-mail: mts@star.ucl.ac.uk; fba@star.ucl.ac.uk; o.lahav@ucl.ac.uk}, F. B. Abdalla$^{1,3}$\footnote[1],, O. Lahav$^{1}$\footnote[1],, D. Kirk$^{1}$, I. Sevilla$^{4}$, \newauthor E. Bertin$^{5}$,  B. T. P. Rowe$^{1}$, J. Annis$^{6}$, M. T. Busha$^{7,8}$, L. N. Da Costa$^{9,10}$, \newauthor J. A. Frieman$^{6, 11, 12}$, E. Gaztanaga$^{13}$, M. Jarvis$^{14}$, H. Lin$^{6}$, W. J. Percival$^{15}$, \newauthor B. X. Santiago$^{10, 16}$, C. G. Sabiu$^{17,18}$, R. H. Wechsler$^{19, 20}$, L. Wolz$^{1,21}$, B. Yanny$^{6}$. 
\\
\\
$^{1}$Department of Physics and Astronomy, University College London, Gower Street, London WC1E6BT, UK \\
$^{2}$Benoziyo Center for Astrophysics, Weizmann Institute of Science, 76100 Rehovot, Israel \\
$^{3}$Department of Physics and Electronics, Rhodes University, PO Box 94, Grahamstown, 6140 South Africa\\
$^{4}$Centro de Investigaciones Energeticas Medioambientales y Tecnologicas, Av.Complutense 40, 28040 Madrid, Spain\\
$^{5}$Institut d'Astrophysique de Paris, UMR 7095 CNRS, Universite Pierre et Marie Curie, 98 bis boulevard Arago, F-75014 \\ Paris, France\\
$^{6}$ Center for Particle Astrophysics, Fermi National Accelerator Laboratory, P.O. Box 500, Batavia, IL 60510, USA\\
$^{7}$ Institute for Theoretical Physics, University of Zurich, Wolfgang-Pauli-Str. 27 CH - 8093 Zurich, Switzerland\\
$^{8}$ Physics Division, Lawrence Berkeley National Laboratory, 1 Cyclotron Road Mailstop 50-4049 Berkeley, CA 94720-8153, USA \\
$^{9}$ Observatorio Nacional, Rua Gal. Jose Cristino 77, Rio de Janeiro, RJ - 20921-400, Brazil \\
$^{10}$ Laboratorio Interinstitucional de e-Astronomia - LineA, Rua Gal. Jos ́e Cristino 77, Rio de Janeiro, RJ - 20921-400, Brazil\\
$^{11}$ Department of Astronomy and Astrophysics, The University of Chicago, 5640 South Ellis Avenue, Chicago, IL 60637, USA  \\
$^{12}$ Kavli Institute for Cosmological Physics, The University of Chicago, 5640 South Ellis Avenue Chicago, IL 60637, USA \\
$^{13}$ Institut de Ciencies de l’Espai (IEEC-CSIC), E-08193 Bellaterra (Barcelona), Spain\\
$^{14}$ Department of Physics and Astronomy, University of Pennsylvania, Philadelphia, PA 19104, USA\\
$^{15}$Institute of Cosmology and Gravitation, University of Portsmouth, Portsmouth, PO1 3FX, UK\\
$^{16}$Departamento de Astronomia, Universidade Federal do Rio Grande do Sul Av. Bento Goncalves 9500, Porto Alegre, \\ RS 91501-970, Brazil\\
$^{17}$Korea Institute for Advanced Study, Dongdaemun-gu, Seoul 130-722,
Republic of Korea \\
$^{18}$Korea Astronomy and Space Science Institute, 776, Daedeokdae-ro, Yuseong-gu, Daejeon, 305-348, Korea\\
$^{19}$Kavli Institute for Particle Astrophysics and Cosmology, SLAC National Accelerator Laboratory, 2575 Sand Hill Rd., \\Menlo Park, CA 94025, USA\\
$^{20}$Physics Department, Stanford University, Stanford, CA 94305, USA\\
$^{21}$Sub-Department of Astrophysics, Department of Physics, University of Oxford, Keble Road, Oxford OX1 3RH, UK}

\begin{document}

\date{Accepted day-- Month Date. Received year month date; in original form year month date}

\pagerange{\pageref{firstpage}--\pageref{lastpage}} \pubyear{2015}

\maketitle
\begin{abstract}
We address the problem of separating stars from galaxies in future large photometric surveys. We focus our analysis on simulations of the Dark Energy Survey (DES). In the first part of the paper, we derive the science requirements on star/galaxy separation, for measurement of the cosmological parameters with the Gravitational Weak Lensing and Large Scale Structure probes. These requirements are dictated by the need to control both the statistical and systematic errors on the cosmological parameters, and by Point Spread Function calibration. We formulate the requirements in terms of the {\it completeness} and {\it purity} provided by a given star/galaxy classifier. In order to achieve these requirements at faint magnitudes, we propose a new method for star/galaxy separation in the second part of the paper. We first use Principal Component Analysis to outline the correlations between the objects parameters and extract from it the most relevant information. We then use the reduced set of parameters as input to an Artificial Neural Network. This multi-parameter approach improves upon purely morphometric classifiers (such as the classifier implemented in {\it SExtractor}), especially at faint magnitudes: it increases the purity by up to 20\% for stars and by up to 12\% for galaxies, at i-magnitude fainter than 23. 
\end{abstract}
\label{firstpage}


\section{Introduction}

What makes a star look different from a galaxy in a deep image? This seemingly very simple question hides the much more complicated issue of allocating a size and a scale to objects observed in the sky, which has concerned observers and theorists throughout the 20th century.
The problem of classifying stars and galaxies in large scale surveys is a long-standing one. It has been encountered back in the early 1990's (e.g. the APM survey,  \citealt{M}) and poses a major challenge for all recent and large imaging cosmological surveys, including the Dark Energy Survey (DES) (http://www.darkenergysurvey.org/) and Euclid (http://sci.esa.int/euclid), which have been designed to uncover the nature of dark energy (DE). One common denominator of the wide variety of observational probes constraining DE is the necessity to select pure samples of galaxies. More specifically, all the surveys must differentiate galaxies at cosmological distances from local objects, to obtain pure, or at least well-understood, samples.

In the area of ``precision cosmology", any source of systematic error is likely to play a decisive role and needs to be taken into account in order to refine the standard inflationary Big Bang picture. An example of a scientific question for which star/galaxy separation is a potentially critical systematic is the precision measurement of Primordial Non-Gaussianities (PNG). These manifest themselves by making the bias of a given type of tracers of dark matter halos strongly scale-dependent. This effect can easily be mimicked by any local systematic effect adding power at large scales and correlated with the galaxies. As the stellar distribution in the Milky Way is across large angular scales, star/galaxy separation is likely to introduce systematic errors in the measurement of PNG. Another example is the effect of occultation of galaxies by stars of comparable magnitudes. \citet{R} showed that this effect constitutes a source of systematic error in the measurement of angular and photometric distributions of luminous red galaxies. Photometric effects associated with faint stars could therefore partially account for the excess power seen in \citet{TAL} for the MegaZ Luminous Red Galaxy survey. This paper gives two other examples, in the case of Weak Lensing (WL) and Large Scale Structures (LSS) measurements, where star/galaxy separation is a key systematic, which needs to be taken into account in order to properly constrain DE.

The outline of this paper is as follows. In section 2, we present the Dark Energy Survey (DES) and the ``DES-like'' simulations which we base our analysis on. In section 3, we study the impact of star/galaxy misclassification on the measurement of the cosmological parameters, in the case of the WL and LSS probes, and show how the requirements on the statistical and systematic errors propagate into new requirements on the quality of star/galaxy separation. In section 4 we summarise the current methods for star/galaxy classification and the motivations for our multi-parameter approach. The details of the method are presented in section 5. In section 6, we compare our star/galaxy classification tool to the ones provided by other methods and confront these results to the science requirements derived in section 3. Finally, we summarise our main conclusions in section 7.

\section{The Dark Energy Survey}

The Dark Energy Survey (DES)\footnote{http://www.darkenergysurvey.org/} is an imaging survey of $5000$ sq-degrees on southern sky, utilising the four meter Blanco telescope in Chile. It will provide imaging of $300$ million galaxies in five filters ({\it g}, {\it r}, {\it i}, {\it z} and {\it Y}). Photometric redshifts will be obtained from the colour information to produce a three dimensional survey. The main goal of DES is to determine the Dark Energy equation of state parameter, $w(z)$, and other key cosmological parameters to high precision. DES will measure $w(z)$ using four complementary techniques in a single survey: counts of galaxy Clusters (GC) (with synergy with clusters detected by the Sunyaev-Zel'dovich effect with the South Pole Telescope), weak gravitational lensing (WL), galaxy power spectra and type Ia Supernovae (SNe). It is expected that the uncertainty on $w(z)$ will be only a few percent for each probe (see \citealt{DES}, for detailed parameterisations and statistics). The science requirements of DES drove the construction of a new camera, the Dark Energy Camera (DECam), which had its first light in September, 2012, and the survey has started in September, 2013.

As part of the process of testing and validation of the DES Data Management (DESDM) system \citep{Mohr}, a series of detailed simulations have been designed to serve as a test-bench for the development of the pipelines and for verifying the scientific reach of the experimental channels. Each of these iterations of the simulations are dubbed ``Data Challenges" (DC). The simulation starts with the creation of galaxy catalogs stemming from an N-body simulation (\citealt{Busha}) and detailed models of the Milky Way galaxy \citep{Rossetto} for the star component. These are merged and fed to an image simulator which includes atmospheric and instrumental effects. The resulting images serve as inputs for DESDM and are processed as the data will be: the code {\it SExtractor} \citep{ba} produces a catalogue of more than 300 parameters encapsulating information about each detected object. 

The most relevant features of these simulations for our study are:

\begin{itemize}
\item the seeing is introduced as a function of observing time;
\item the galaxy shapes have been implemented using a Sersic profile which matches the observed profile;
\item the Point Spread Function (PSF) takes into consideration the seeing for that time, the optics and the distortion as a function of separation from the optical axis.
\end{itemize}

The results shown in this paper are based on the latest release (internal to the DES collaboration) of simulated data, DC6, which covers approximately 140 square degrees to the full DES depth, corresponding to about 10 nights of observations. 
We select from it the objects with a model magnitude in the {\it i} band brighter than $24$, as they are the ones most likely to be detected with DES. 

\section{science requirements on star/galaxy separation}

DES will be among the first surveys to combine in a single project the observation of the four preferred dark energy probes, as identified by the Dark Energy Task Force (DETF) \citep{Alb}. SNe and Baryonic Acoustic Oscillation (BAO) constrain the expansion of the Universe as a whole and are therefore referred to as {\it purely geometric}. WL and GC constrain both the expansion on the Universe and the growth of Large Scale Structures (LSS) (See  \citealt{Wein} for a complete review). 

In order to properly constrain DE, the broad variety of measures carried out within each probe must meet certain requirements defined by DES science teams. While there is no unique way to specify the constraints on dark energy experiments and probes, the Figure of Merit (FoM), defined by the DETF, provides a useful metric. If we parameterise the time evolution of DE by the equation of state $w(a)=w_o+(1-a)w_a$, where $a(t)=\frac{1}{1+z(t)}$ is the cosmic scale factor and $z(t)$ is the redshift of an object emitting at time t, the FoM is defined as the reciprocal of the area of the error ellipse enclosing $95 \%$ confidence limit in the $w_o$-$w_a$ plane. Larger FoM indicates smaller errors and therefore greater accuracy on the measurement of the parameters. 

In other words, reaching the FoM goals requires to minimise the error on $w_o$ and $w_a$.
Since the total error is the sum of the {\it statistical} error and the {\it systematic} error
, we can derive two types of science requirements. More concretely, the total Mean Square Error (MSE) on a cosmological parameter $p_\alpha$ can be decomposed as
\begin{equation}\label{eq:MSE}
MSE[p_{\alpha}]=\sigma^2[p_{\alpha}]+\Delta^2[p_\alpha]\;,
\end{equation}
where $\sigma^2[p_{\alpha}]$ is the statistical error variance and $\Delta[p_\alpha]$ is the parameter shift due to the systematic signals.
For each probe, both of these terms needs to be controlled in order to minimise the total error.

Star/galaxy misclassification is an interesting effect because it contributes to both the statistical and systematic part of the total error, for the WL and LSS probes. This allows us to translate separately the requirement on the statistical term (section $3.2$) and the requirements on the systematic term (section 3.3) into requirements on the quality of the star/galaxy separation. Additional requirements are specific to each probe, e.g. PSF calibration for WL (section $3.4$).

We outline below a formalism to derive these requirements.

\subsection{Formalism}

\subsubsection{Completeness, contamination and purity}

In the following, we define the parameters used to quantify the quality of a star/galaxy classifier. For a given class of objects, $X$ (stars or galaxies), we distinguish the surface density of well classified objects, $N_X$, and the density of misclassified objects, $M_X$. 
 \\
 \\
  \begin{tabular}{|| l || c || r ||}
  \hline
& True Galaxies & True stars \\
\hline
Objects classified as galaxies & $N_G$ & $M_S$ \\
\hline
Objects classified as stars & $M_G$ & $N_S$\\
 \hline

\end{tabular}\\ \\

The galaxy {\it completeness} $c^g$ is defined as the ratio of the number of true galaxies classified as galaxies to the total number of true galaxies. The stellar contamination $f_s$ is defined as the ratio of stars  classified as galaxies to the total amount of objects classified as galaxies. 

\begin{equation}\label{eq:cg}
c^g=\frac{N_G}{N_G+M_G} \;,
\end{equation}

\begin{equation}\label{eq:fs}
f_s=\frac{M_S}{N_G+M_S}\; .
\end{equation}
The {\it purity} $p^g$ is defined as $1-f_s$:
 \begin{equation}\label{eq:pg}
 p^g=\frac{N_G}{N_G+M_S}=1-f_s\;.
 \end{equation} 
Similar parameters can be defined for a sample of stars: $p^s$, $f_g$ and $c^s$.

We aim to formulate the requirements on the statistical and systematic errors in terms of constraints on these parameters. This will allow us to quickly compare the performance of the classifiers presented in section $4$ and $5$ and assess whether they allow us to achieve the goals of the DETF FoM.

One should note that there are some inefficiencies in the image pipeline, which are studied in DC6 and which we do not deal with in this analysis. Instead, we define the latter parameters with respect to the mock galaxy samples used to produce the image simulations. With real DES data, our results could be tested e.g. on HST data in the same fields. 

\subsubsection{Fisher Information Matrix}

The Fisher information matrix describes how the errors on the angular power spectrum $C(l)$ (of the cosmic shear in the case of WL, and the density fluctuations of galaxies in the case of LSS) propagate into the precision on the cosmological parameters $p_{\alpha}$ . We employ this formalism (see \citealt{Teg}, for a review), to quantify the impact of star/galaxy misclassification on each of the terms in equation~\ref{eq:MSE}, i.e. on the statistical and systematic errors on the cosmological parameters.

The Fisher matrix can be expressed as
\begin{equation}\label{eq:F}
F_{\alpha \beta} = \sum\limits_l \sum\limits_{(i,j)(m,n)}  \frac{\partial C_{ij}(l)}{\partial p_\alpha}Cov^{-1}[C_{ij}(l),C_{mn}(l)]\frac{\partial C_{mn}(l)}{\partial p_\beta} \; ,
\end{equation}
where the sum is over multipole values and redshift bins (typically five for WL). $Cov[X,Y]$ designates the covariance matrix of $X$ and $Y$ and is given by \citep{Takada},
\begin{equation}\label{eq:Cov}
Cov[C_{ij}(l),C_{mn}(l)]=\frac{ \{ C_{im}(l)C_{jn}(l)+C_{in}(l)C_{jm}(l) \}}{f_{sky}(2l+1)\Delta l} \;,
\end{equation}
where $f_{sky}$ is the fraction of the sky covered by the survey ($f_{sky}=0.1212$ for DES) and $\Delta l$ is the width of the corresponding angular frequency bin. 

\subsection{Science requirements on the statistical errors}

How does the need to control the statistical errors on the cosmological parameters propagate into a requirement on the quality of star/galaxy separation? In the following, we aim to answer this question in the case of the WL and LSS probes.

\subsubsection{WL measurements}

Gravitational lensing from distant intervening mass fluctuations causes the shapes of objects to be distorted such that they appear to be more or less elliptical. While no single object is intrinsically round, if the intrinsic shapes of galaxies are uncorrelated with one another, one can average the apparent shapes of many thousands of such objects to extract a distortion attributed to WL. The statistical properties of this observable pattern put a constraint on the power spectrum and therefore on the cosmological model and on DE. For some concise introductions to cosmic shear, see e.g. \citet{Mellier}, \citet{Bart} and \citet{Refri}. 

How do star/galaxy misclassifications affect the WL shear measurement? The predicted shear angular power spectrum $C_{ij}(l)$ depends on $N_{eff}$, the effective density per unit area of galaxies with reliable shape measurements,

\begin{equation}\label{eq:Cl}
C_{ij}(l) =\int_{0}^{r_H}drr^{−2}W_i(r)W_j(r)P(l/r;r)+\delta_{ij}\frac{\sigma_e^2}{N_{eff}}
\end{equation}
where $P(k = l/r)$ is the $3$D matter power spectrum, $W_i(r)$ and $W_j(r)$ are the radial window functions of the redshift bins $(i, j)$, $r$ is the comoving distance and $r_H$ is the Universe horizon. The angular power spectrum depends on $N_{eff}$ through the last term, i.e. the ``shot noise'' due to $\sigma_e$, the intrinsic ellipticity noise for the galaxy sample. 

In order study the effect of $N_{eff}$ on the statistical error $\sigma[p_\alpha]$,  we compute the Fisher matrix for different values of $N_{eff}$. We estimate the $C_{ij}(l)$ and $\frac{\partial C_{mn}^{g}(l)}{\partial p_\alpha}$ terms (see Eq.~\ref{eq:F}) using the same code as in \citet{Lazlo} and \citet{Kirk2}. The setup is as follows: we use a model with eight free parameters: \{$w_o$, $w_a$, $\Omega_m$, $H$, $\sigma_8$, $\Omega_b$, $n_s$, $b_g$\}; we assume a Planck prior (Jochen Weller, personal communication); there are five tomographic bins of roughly equal number density between z = 0 and 3; the redshift distribution is a Smail-type distribution (e.g. equation (12) of \citealt{Amara}, with $\alpha=2$,  $\beta=1.5$, $z_0 = \frac{0.8}{1.412}$); we compute the $C_{ij}(l)$ and $\frac{\partial C_{mn}^{g}(l)}{\partial p_\alpha}$ terms for $l\in [1,1024]$, to avoid the  strongly non linear regime where baryon physics will start being important and following the l-cuts performed in most recent works by the WL comunity \citep{DAS,DEB,AUD}; and the photometric redshift error is $\Delta z = 0.05*(1+z)$.

We then compute the marginalized statistical error on the cosmological parameters by approximating them with their Cramer-Rao lower bound
\begin{equation}
\sigma[p_\alpha] \approx \sqrt{(F^{-1})_{\alpha \alpha}}
\end{equation}
We show the results for $w_o$ and $w_a$ in figure~\ref{fig:neff_sigma_WL} and for the other free parameters of our model in the Appendix.

Figure~\ref{fig:neff_sigma_WL} shows that larger $N_{eff}$ translates into smaller statistical errors on $w_o$ and $w_a$, i.e. larger FoM, which puts a constraint on $N_{eff}$: it has to be higher than a threshold value $N_{thresh}$ which can depend on the bandpass considered, 

\begin{equation}\label{eq:Neff1}
N_{eff} \geq N_{thresh} \; .
\end{equation}
Figure~\ref{fig:neff_sigma_WL} also shows asymptotes above $N_{thresh}=10$, i.e. the effect of any variation of $N_{eff}$ on the statistical error decreases at high $N_{eff}$. In practice, we require the increase of the statistical error due to star/galaxy missclassification to be smaller than $2\%$. If this reasonable but somewhat arbitrary goal is not achieved, it will only increase the statistical error and will not lead to a bias of the WL results. This translates into a decrease of $N_{eff}$ smaller than $4\%$, i.e. 
\begin{equation}\label{eq:cg2}
c^g \ge  96.0\%
\end{equation}
Star-galaxy misclassification is only one among many other sources of errors leading true galaxies to be rejected from the sample of galaxies with reliable shape measurements, (e.g., shape measurement errors and photo-Z errors). To insure that the statistical errors are controled, this condition on $c^g$ should be completed by constraints on the survey parameters controling all the other sources of errors.  
\begin{figure}
\includegraphics[width=8cm]{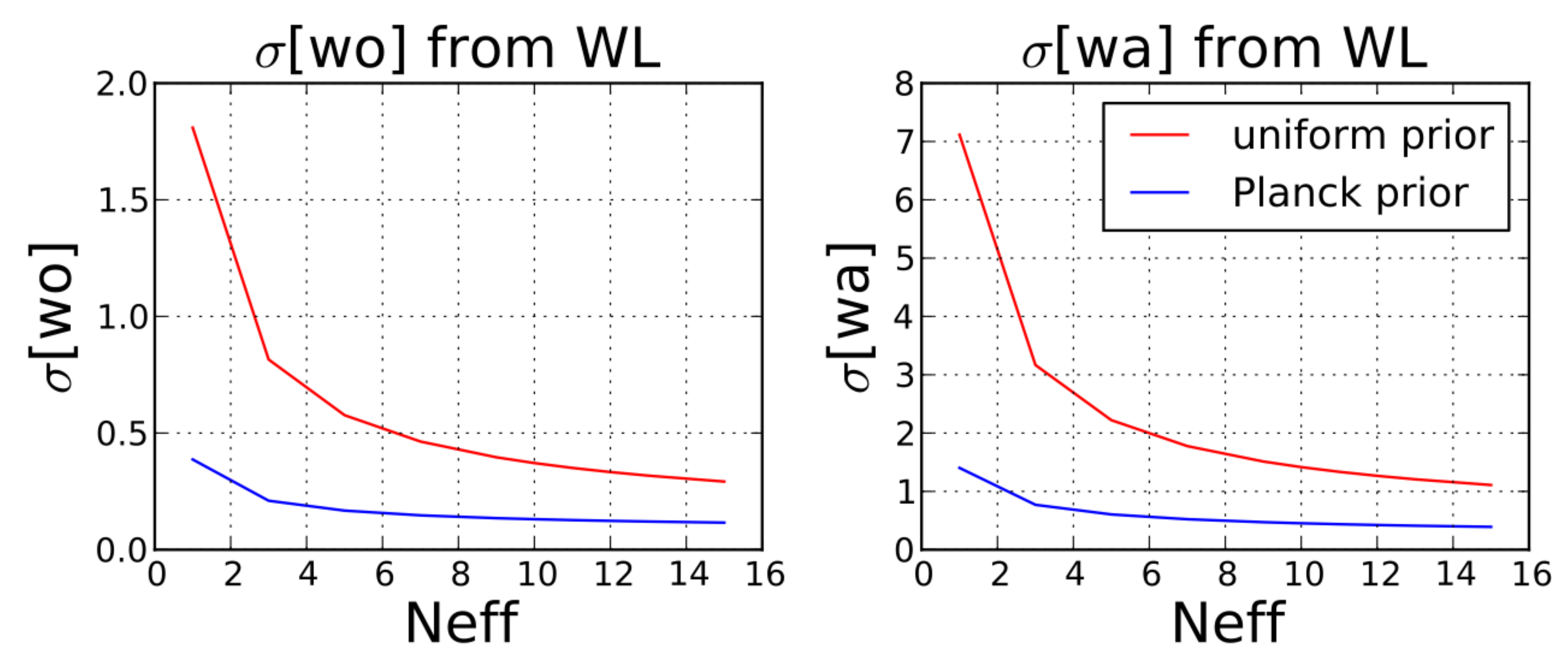}
\caption{Marginalised statistical errors on the equation of state parameters $w_o$ and $w_a$ from the WL probe, for different values of the density of galaxies with reliable shape measurement $N_{eff}$. The errors are marginalised over \{$\Omega_m$, $H$, $\sigma_8$, $\Omega_b$, $n_s$, $b_g$\} and computed  using the assumptions and setup described in section $3.2.1$. The red curve shows the errors computed with a non-informative prior whereas the blue curve is obtained assuming a Planck prior.} 
\label{fig:neff_sigma_WL}
\end{figure}

\subsubsection{LSS measurements}
LSS measurements allow us to constrain DE in various ways. The position of the BAO feature provides a standard ruler to study the expansion history. The shape of the angular power spectrum of the galaxy density fluctuation encapsulates precious information about the clustering amplitude and the growth of structures. 

\begin{figure}
\includegraphics[width=8cm]{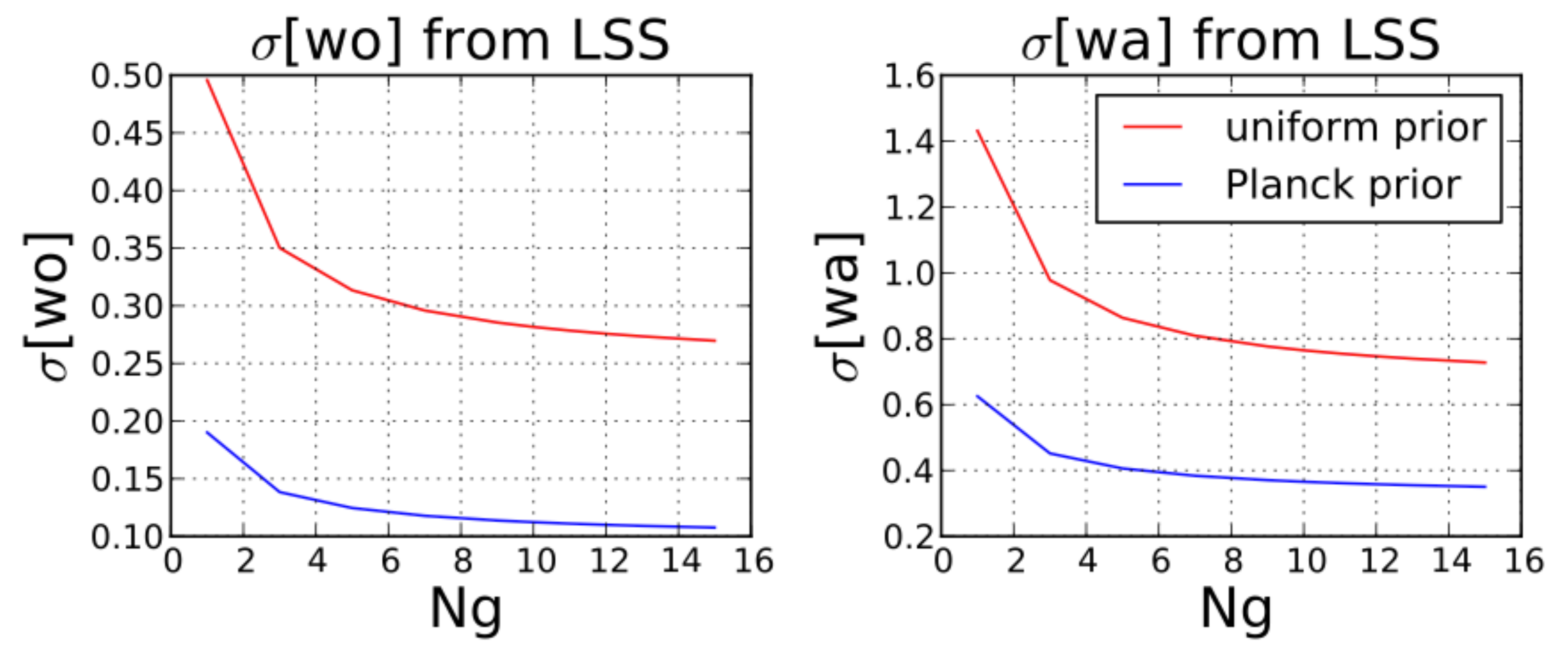}
\caption{Marginalised statistical errors on the equation of state parameters $w_o$ and $w_a$ from the LSS probe, for different values of the density of detected galaxies $N_g$. The errors are marginalised over \{$\Omega_m$, $H$, $\sigma_8$, $\Omega_b$, $n_s$, $b_g$\} and computed using the same assumptions and setup as in the WL case (see section $3.2.1$), with $l\in [10, 400]$, to avoid the non linear regime and following most recent l-cuts work by the LSS community \citep{Ras,AUD}. The red curve shows the errors computed with a non-informative prior whereas the blue curve is obtained assuming a Planck prior.}
\label{fig:neff_sigma_LSS}
\end{figure}

Star/galaxy misclassification affect the power spectrum measurements and the statistical error on the cosmological parameters in  a similar way as in the WL case. Indeed, we can write the same equation as Eq.~\ref{eq:Cl} for the angular power spectrum of the galaxy density fluctuations. The shot noise term is then given by $\frac{1}{N_G}$, where $N_G$ is simply the surface density of detected galaxies. In figure~\ref{fig:neff_sigma_LSS}, we show the evolution of the statistical errors on $w_o$ and $w_a$ with the density of detected galaxies, computed using the same setup as in the WL case.

During the design phase of the project, it has been estimated that in order to achieve the goals of the the LSS FoM, the 5000 sq-degrees DES survey will need to provide reliable photo-z and position measurement for about 200 millions galaxies, i.e. the number of galaxies correctly classified $N_G$ should be higher than $11.1$ per sq-arcminute (when using combined measurements from the {\it r}, {\it i} and {\it z} bandpasses). This requirement is currently being re-visited using data. When doing the latter calculation on the truth table of DC6, for which the surface density of galaxies is $N^g_{tot}\approx 12.5$, this threshold on $N_G$ translates into the following requirement on the galaxy completeness provided by the star/galaxy classifier: $c^g > 88.9\%$.

Note that galaxies at different redshifts have different weights. This is somewhat related to the magnitude of galaxies, as the distribution of brighter galaxies will peak at a lower redshift than galaxies which are fainter. Even though this is not explicitly stated when mentioning that there is a requirement of 200 millions galaxies, we are implying that we are effectively going to a given depth and therefore sampling galaxies out to a given redshift. These caveat are implicitly included within the Fisher matrix calculation.

Note also that the derived requirement is a necessary but not sufficient condition, as other sources of errors, apart from star/galaxy misclassification (e.g. photo-z errors), reduce the number of galaxies which can be used for LSS measurement.
\subsection{Science requirements on the systematic errors}

We now explore the contribution of star/galaxy misclassification as a source of {\it systematic} error, which need to be controlled in order for the FoM objectives to be achieved. Star/galaxy misclassifications generate a residual signal $\delta C^{sys}(l)$ in the angular power spectra (of the cosmic shear in the case of WL, and the density fluctuations of galaxies in the case of LSS), which propagates into a systematic shift $\Delta[p_\alpha]$ of the cosmological parameter $p_\alpha$.
We use the same formalism as in \citet{Amara} (see also \citealt{Kirk} and \citealt*{Huterer}), to derive $\Delta[p_\alpha]$,
\begin{multline}\label{eq:B}
\Delta[p_\alpha]=\\
\sum\limits_{\beta,l,(i,j),(m,n)} (F^{-1})_{\alpha \beta}\delta C_{ij}^{sys}(l)Cov^{-1}[C^{gal}_{ij}(l),C^{gal}_{mn}(l)]\frac{\partial C^{gal}_{mn}(l)}{\partial p_\beta} \;,
\end{multline}
where $F^{-1}$ is the inverse Fisher matrix. A criterion usually used to constrain the contribution of the systematic error to the total MSE, is to define a tolerance on the systematics such that they do not dominate over statistical error. This is verified when

\begin{equation}\label{eq:sum}
\left|\Delta[p_\alpha] \right| \leq \sigma[p_\alpha] \; ,
\end{equation}

In the following sections, we derive the systematic parameter shift for 7 cosmological parameters $p_\alpha=\left\{w_o, w_a, \Omega_m, H, \sigma_8, \Omega_b,n_s\right\}$ and the galaxy bias $b_g$, in the case of WL and LSS. This allows us to translate Eq.~\ref{eq:sum} into requirements on the quality of the star/galaxy separation.

\subsubsection{Requirement from WL measurements}

In the case of WL, the systematic error $\delta C^{sys}(l)$ comes from the fact that some stars are identified as galaxies, and therefore contribute to the measured cosmic shear.
We decompose the measured shear $\gamma_m$ into the contribution from the true galaxies and the contamination from the misclassified stars. The galaxy shear is measured by deconvolving the observed shear and a PSF model, therefore the contamination from stars in a galaxy sample appears as a residual deconvolved shear:

\begin{equation}\label{eq:gamma_m}
\gamma _{m}=(1-f_s)\gamma _{g}+f_s\gamma _{s,res} \;.
\end{equation}
where $f_s=1-p^g$, is the stellar contamination rate (defined in Eq.~\ref{eq:fs}) and $\gamma_{s,res}$ is the residual PSF shear, after deconvolution of the PSF model from the shape of misclassified stars. In the following analysis, we make a toy model where the residual deconvolved shear can be written as
\begin{equation}
\gamma_{s,res}=a\gamma_s\;,
\end{equation}
where $a \in [0,1]$ and $\gamma_s$ is the stellar shear. The measured two-point shear correlation function is then
\begin{equation}\label{eq:gamma_m2}
 <\gamma_m\gamma_m>=(1-f_s)^2<\gamma _g\gamma_g>+f_s^2\alpha<\gamma _s\gamma_s> \;,
\end{equation}
and in terms of measured angular power spectrum, the latter equation reads
\begin{equation}\label{eq:cobs2}
C^{obs}(l)=(1-f_s)^2C^{gal}(l)+f_s^2 \alpha C^{s}(l)\;,
\end{equation} 
where $\alpha=a^2$ and where we assumed that $\gamma_g$ and $\gamma_s$ are uncorrelated. In practice, this is not necessarily the case. Our toy model introduces into the same term, $\alpha C_s(l)$, the auto-correlation of  the residual ``deconvolved star shapes'' and possible cross-correlation between them and the galaxy shear $\gamma_g$. Setting $\alpha$ to zero comes to neglecting both of these terms, and setting $\alpha=1$ comes to overestimating them both.  We derive the requirement on the quality of star/galaxy separation in the two limiting cases $\alpha=1$ and $\alpha=0$ and leave the more general case for further analysis.
Equation~\ref{eq:cobs2} gives the residual systematic signal
\begin{equation}\label{eq:deltaC}
\delta C^{sys}(l)=f_s^2(C^{gal}(l)+\alpha C^{s}(l)) -2f_s C^{gal}(l) \;.
\end{equation}
The requirement stated in Eq.~\ref{eq:sum} can be reformulated as a requirement on the stellar contamination rate $f_s$,
\begin{equation}\label{eq:pfs}
\mathcal{P}(f_s) \leq 0 \;,
\end{equation}
where $\mathcal{P}$ is a second order polynomial.

The assumptions made to solve Eq.~\ref{eq:pfs} are detailed below.
We use the setup detailed in section $3.2.1$ to compute the Fisher matrix and the marginalised statistical errors $\sigma[p_\alpha]$ on the cosmological parameters. 
To estimate $C^s(l)$ in Eq.~\ref{eq:deltaC}, we assume it is the sum of a ``shot noise'' term and a term due to the correlation of stellar shapes across the field of view,
\begin{equation}\label{eq:tile}
C^s(l)=C^s_{noise}+C^s_{tile}(l)
\end{equation}
We measure $C^s_{tile}(l)$, the power spectrum of the shapes of the stars in DC6, using the same code as in \citet{JB}. The ``shot noise'' term is given by
\begin{equation}\label{eq:Cls}
C^s_{noise}= \frac{\sigma_s^2}{N^{s}_{tot}}
\end{equation}
where $N^{s}_{tot}=N_s+M_s$ (see section $3.1.1$) is the density of stars and $\sigma_s$ is the ellipticity of stars. Various complex effects combine as different terms in the stellar ellipticity. First, the interpolated PSF can only constrain large-scale modes of the PSF variation. The power on small scales is not well-constrained at all, so it will show up more or less completely in the shapes of stars that are corrected by the PSF model. Second, the deconvolution process magnifies the errors in the shape measurement, especially for objects that are nearly the same size as the PSF, which is presumably the case for stars that are misidentified as galaxies. We approximate the contamination $\sigma_s$ as being just the original PSF ellipticity. We believe this is a reasonably conservative treatment given the complexity of the effects that combine as different terms in this ellipticity.

To estimate $\sigma_s$, we use the {\it whisker length}. Given $I_{xx}$, $I_{yy}$ and $I_{xy}$, the second moment of the light intensity from an object in $x,y$ coordinates, a measure of the ellipticity of the light distribution is given by $e=(I_{xx}-I_{yy})(I_{xx}+I_{yy})$. The whisker length is then defined as $w \approx \sqrt{e(I_{xx}+I_{yy})}=\sqrt{e} \cdot r_{psf}$, where $r^2_{psf}$ is given by $(FWHM)/2.35$. $FWHM$ designates the full width at half maximum and is given by $FWHM \approx 0.94$ in DES. In addition, the hardware has been designed with a requirement on the whisker length to be lower than a threshold value of 0.2'' in the {\it r}, {\it i} and {\it z} band, which we take as an estimation of $whisk$. We get $C^s\approx 1.3187\cdot10^{-8}$ sr. 

\begin{figure*}
\begin{center}
\includegraphics[width=14cm]{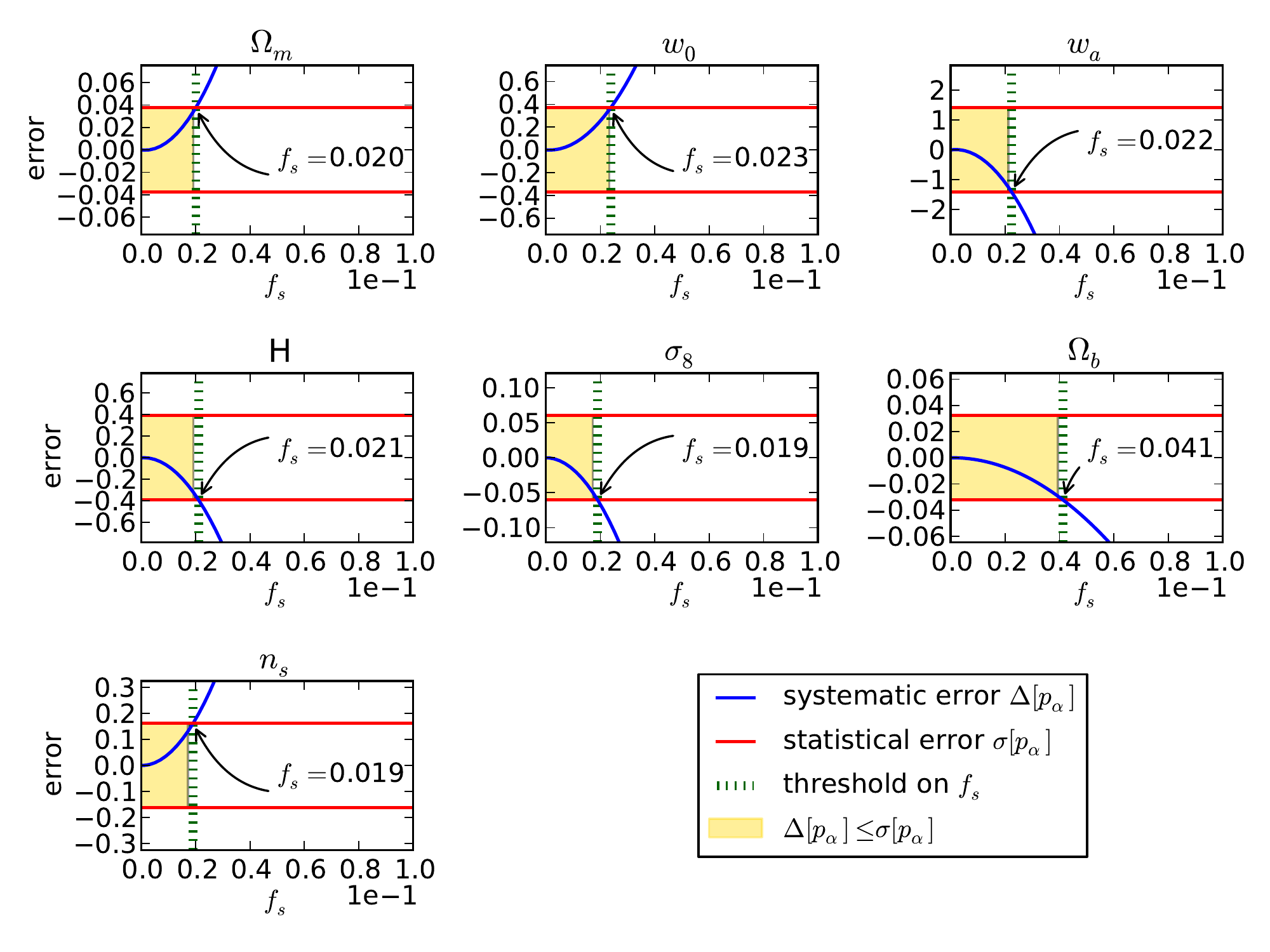}
\caption{Marginalised statistical error $\sigma$ (red line) and systematic parameter shift $\Delta$ (blue curve) from the WL probe, for different values of the stellar contamination $f_s$ allowed by the star/galaxy classifier. Both $\sigma$ and $\Delta$ are marginalised over \{$\Omega_m$, $H$, $\sigma_8$, $\Omega_b$, $n_s$, $b_g$\} and are computed using the setup described in section $3.2.1$. The yellow area corresponds to the values of $f_s$ for which the requirement on the systematic errors is achieved, i.e. it does not dominate over the statistical error. This requirement translates into a threshold on $f_s$, indicated by the green line. Unlike LSS measurements, WL measurements are not sensitive to the galaxy bias $b_g$, which is the reason why it does not appear above.}
\label{fig:plot_constrains_WL}
\end{center}
\end{figure*}

\begin{figure}
\begin{center}
\includegraphics[width=7cm]{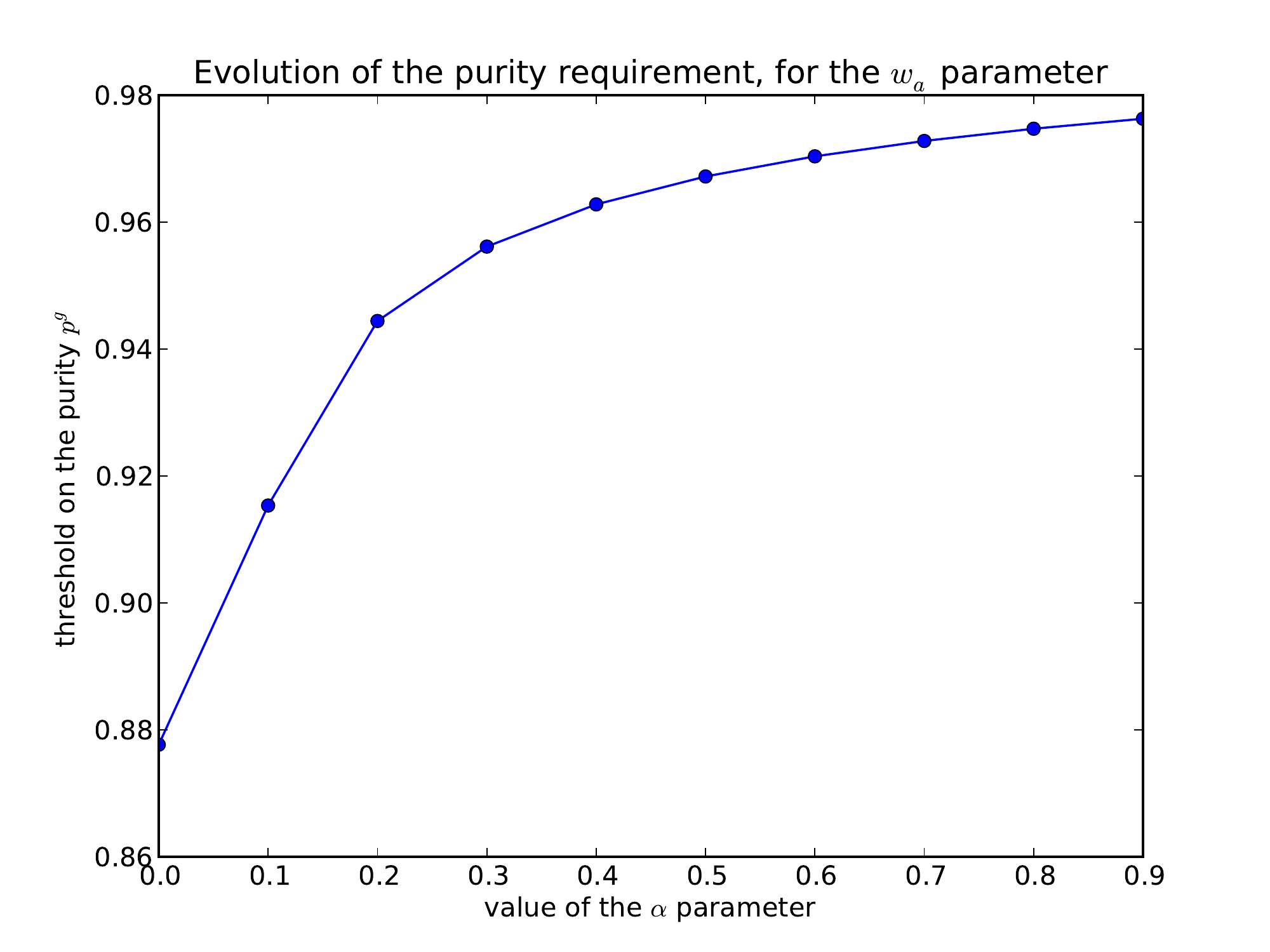}
\includegraphics[width=7cm]{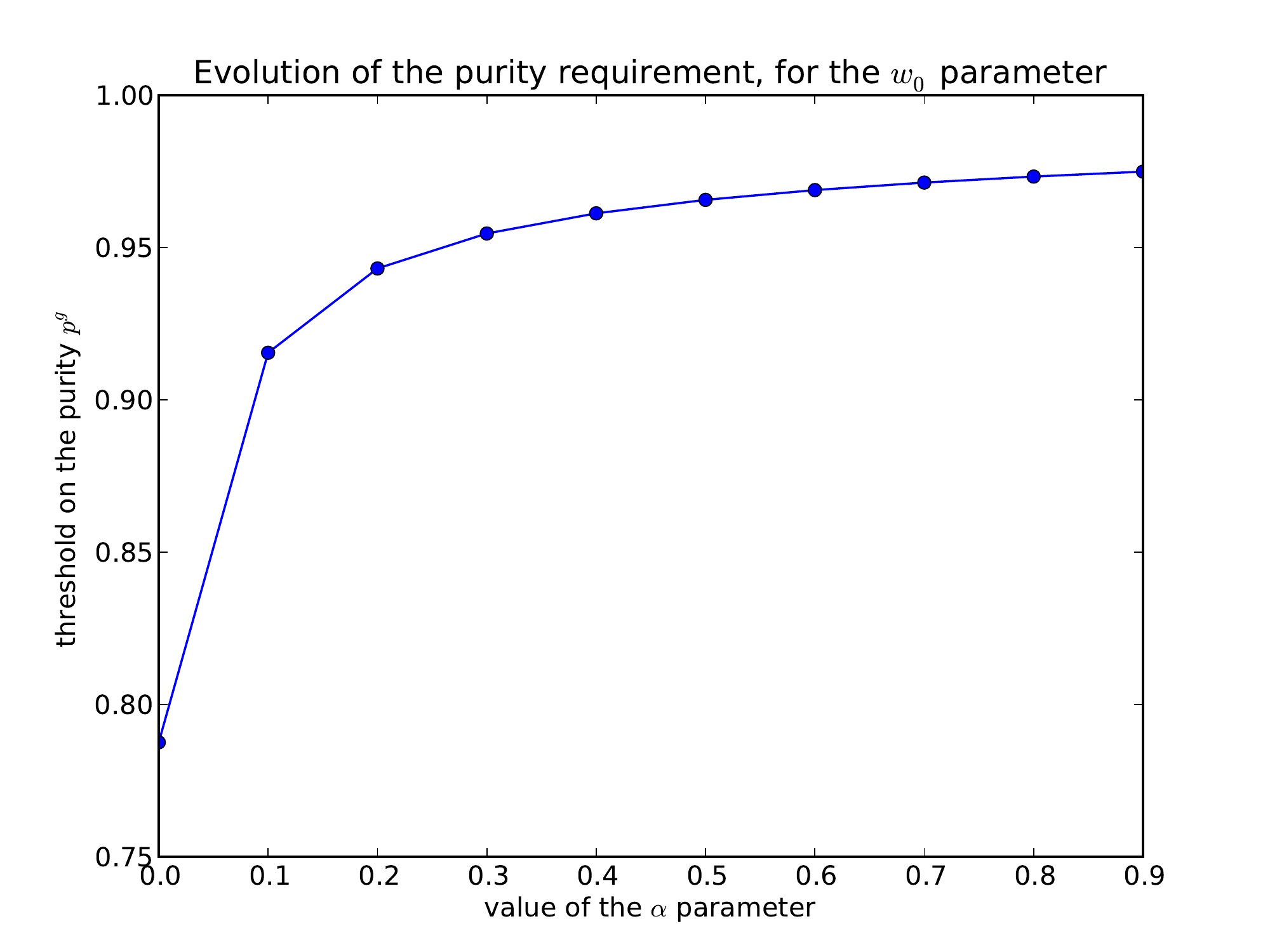}
\caption{Evolution with the coefficient $\alpha$ of the value of $p^g_{lim}$, from the constraint on the bias of the equation of star parameter $w_a$ (top) and $w_o$ (bottom).}
\label{fig:plot_constrains_wawo}
\end{center}
\end{figure}

Here we consider the two limiting cases $\alpha=0$ and $\alpha=1$  and derive the  lower bounds for $f_s$ corresponding to each of these cases, referred to as $f_{s,lim,\alpha=0}$ and $f_{s,lim,\alpha=1}$. The true lower bound is in the interval corresponding to these limiting cases:  $f_{s,lim}\in[f_{s,lim,\alpha=1}, f_{s,lim,\alpha=0}]$. In particular, Figure~\ref{fig:plot_constrains_WL} shows the limiting case $\alpha=1$: we plot the two terms of the total error $MSE[p_\alpha]$ (see equation~\ref{eq:MSE}), i.e. the systematic parameter shift $\Delta[p_\alpha]$ due to star/galaxy misclassification, and the statistical error $\sigma[p_\alpha]$, for different values of the stellar contamination $f_s$ and for each of the cosmological parameters of our model $p_\alpha=\{w_o, w_a, \Omega_m, H, \sigma_8, \Omega_b, n_s, b_g\}$. For the equation of state parameters $w_o$ and $w_a$, we find that we require $f_s \leq f_{s,lim}$ with $f_{s,lim,\alpha=0}=0.122$ and $f_{s,lim,\alpha=1}=0.022$ (requirement driven by $w_a$).  This translates into the following requirement on  $p^g=1-f_s$, the purity provided by the star/galaxy classifier: $p^g\ge p^g_{lim}$ with $p^g_{lim} \in [87.7\%,97.8\%]$. To refine this requirement, we now allow $\alpha$ to vary. In figure~\ref{fig:plot_constrains_wawo}, we show the evolution of $p^g_{lim}$ when varying $\alpha$ and when considering the requirement on the parameters $w_o$ and $w_a$. The threshold is driven by $w_a$ (since the requirement to constrain the bias on $w_a$ leads to a more stringent value of $p^g_{lim}$). The value of $p^g_{lim}$ quickly grows with $\alpha$. Above $\alpha=0.4$, $p^g_{lim}$ grows slower and stays above 96\%.

Within an experiment designed to constrain DE such as DES, the constraints on the quality of star/galaxy separation comes from the need to control the errors on $w_o$ and $w_a$. This being said, one should keep in mind that the contamination from stars affects the precision on the measurements of other cosmological parameters, as shown in figure~\ref{fig:plot_constrains_WL}. 

\subsubsection{Requirement from LSS measurement}

Achieving the objectives of the LSS FoM requires the systematic error induced by star/galaxy misclassification to be smaller than the statistical error on $w_o$ and $w_a$, and we can rewrite Eq.~\ref{eq:sum} in the case of LSS measurements. The shape of the residual systematic signal due to star/galaxy misclassification, $\delta C^{sys}$, is obtained following the same methodology as in the WL case, by decomposing the measured density fluctuation into the contribution from the true galaxies and the contamination from the stars identified as galaxies,

\begin{equation}\label{eq:delta_m}
\delta _{m}=(1-f_s)\delta _{g}+f_s\delta _{s} \;. 
\end{equation}

Replacing the shear angular power spectrum with the density fluctuation angular power spectrum in Eq.~\ref{eq:deltaC}, we get the same requirement on the stellar contamination rate $f_s$ as in Eq.~\ref{eq:pfs}. To estimate $C^s(l)$, we use the same stellar catalogue as used for the DES simulated sky survey produced by \citet{Busha}. We then calculate $C^s (l)$ using the approach from \citet{TAL2} and an adaptation of the HEALPix code \citep{Gorski}. We estimate the $C_{ij}(l)$ and $\frac{\partial C_{mn}^{g}(l)}{\partial p_\alpha}$ terms using the same code and setup as for the WL case. 
\begin{figure*}
\begin{center}
\includegraphics[width=14cm]{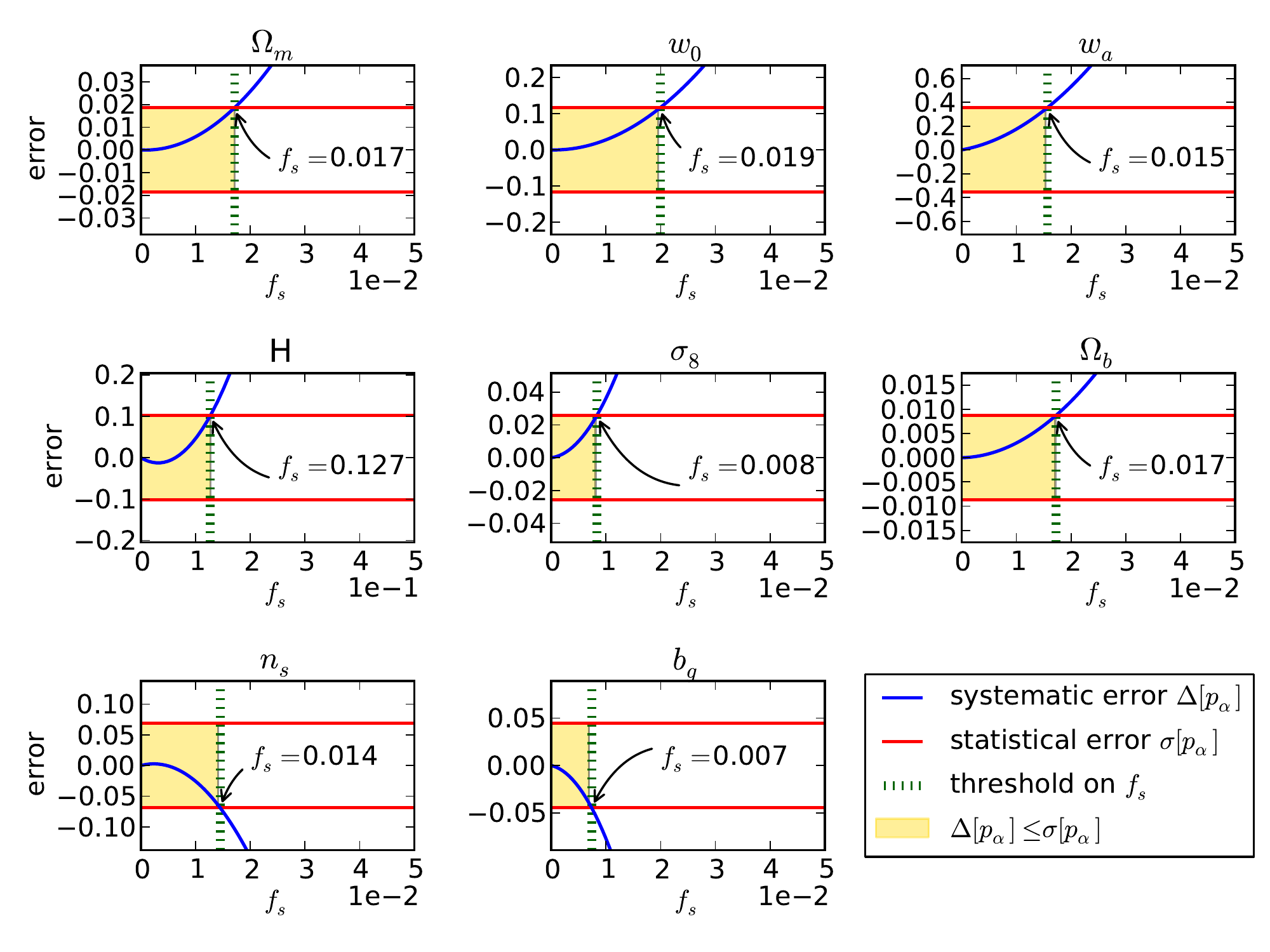}
\caption{Marginalised statistical error $\sigma$ (red line) and systematic parameter shift $\Delta$ (blue curve) from the LSS probe, for different values of the stellar contamination $f_s$ allowed by the star/galaxy classifier. Both $\sigma$ and $\Delta$ are marginalised over \{$\Omega_m$, $H$, $\sigma_8$, $\Omega_b$, $n_s$, $b_g$\} and are computed using the setup described in section $3.2.1$, with $l \in [10, 400]$, to avoid the non linear regime. The yellow area corresponds to the values of $f_s$ for which the requirement on the systematic errors is achieved, i.e. it does not dominate over the statistical error. This requirement translates into a threshold on $f_s$, indicated by the green line. Unlike WL measurements, LSS measurements are sensitive to the galaxy bias $b_g$, as shown on the last panel.}
\label{fig:plot_constrains_LSS}
\end{center}
\end{figure*}
Figure~\ref{fig:plot_constrains_LSS} shows the systematic parameter shift induced by the stellar contamination,  for each of the cosmological parameters of our model $p_\alpha=\{w_o, w_a, \Omega_m, H, \sigma_8, \Omega_b, n_s, b_g\}$. In particular, for the equation of state parameters $w_o$ and $w_a$, we find that we require $f_s \leq 0.015$. This translates into the following requirement on $p^g=1-f_s$, the purity provided by the star/galaxy classifier: $p^g \ge 98.5\%$.
The requirement on star/galaxy separation in a DE experiment is dictated by the need to accurately measure $w_o$ and $w_a$. This being said, figure~\ref{fig:plot_constrains_LSS} demonstrates that these two parameters are not the most sensitive to the contamination by stars, which we leave for further analysis. 

\subsection{Stellar PSF calibration for WL}

In this section, we derive two additional requirements on the quality of the star/galaxy separation, from calibration constraints specific to the WL probe. The measured shapes of galaxies include a component due to the PSF of the combined telescope, atmosphere, and instrument which is correlated among galaxies. Removing this contribution requires careful measurement of the PSF, which is done using isolated stars. Therefore, additional requirements on star/galaxy separation come from PSF calibration for WL.

\subsubsection{Requirement on $c^s$}

In order to determine the interpolation pattern of the PSF, one needs to find enough stars to adequately cover the area of the CCD chip. Based on preliminary studies of the DES science verification data, we believe around $200$ stars per DES CCD is enough to adequately cover the area of the CCD chip and determine the interpolation pattern of the PSF. From the truth tables, we know that the total number of stars per CCD is approximately 810 and therefore the technical constraint on the completeness of the stars samples is $c^s \ge 25\%$.

In this analysis, we assumed that all non-saturated stars can be used for PSF estimation. In practice, the latter lower limit on the completeness could be more stringent because of detector non-linearities. Indeed, the ``blooming'' effect, caused by the voltages induced by the photons reaching the detector, leads brighter objects to appear larger than faint objects. This effect can lead to variations of the PSF between bright and faint stars, and therefore affect the PSF calibration. This reduces the number of stars available for PSF calibration. 

\subsubsection{Requirement on $p^s$}

The upper limit on the contamination in a sample of stars comes from the fact that galaxies misclassified as stars will bias the inferred PSF, which in turn will bias the galaxy shapes.  
We use a toy model to estimate the bias on the shear estimate as a function of $f_g=M_g/(N_s+M_g)$, the galaxy contamination rate in the sample of stars.

Let us first consider the sample of objects classified as stars, used for the calibration of the PSF. Such a sample actually includes two types of objects:  true stars and true galaxies which have been misclassified as stars. The PSF model derived from this sample can be approximated as the weighted average of both types of objects:
\begin{equation}\label{eq:chibiasx}
\chi_{psf}^{biased}(f_g)=f_g\chi^{mis,gal}+(1-f_g)\chi_{psf}^{true}\;,
\end{equation}
where $\chi$ is the polarisation, and is related to the observed major and minor axis $a$ and $b$ of the image produced by a circular source via
\begin{equation}
|\chi|=\frac{a^2-b^2}{a^2+b^2}\;,
\end{equation}
and to the shear and convergence fields via
\begin{equation}
\chi=\frac{2\gamma(1-\kappa)}{(1-\kappa)^2+|\gamma|^2}\;,
\end{equation}
so that $|\chi|\approx 2|\gamma|$.

Let us now consider a sample of galaxies of which we would like to measure the shear. 
The observed polarisation $\chi^{obs}$, i.e. the polarisation after convolution with the PSF model, is linked to the true polarisation of a galaxy through the following relation \citep{Viola}:
\begin{equation}
\chi^{obs}_{gal}=\frac{\chi_{gal}^{true}}{1+1/R}+\frac{\chi_{psf}^{true}}{1+R}\;.
\end{equation}
The {\it resolution} $R$ in the above equation is the ratio of the galaxy to PSF size. 
In the absence of misclassified galaxy contaminating the sample used to measure $\chi_{psf}$ (and neglecting the other sources of errors in the PSF calibration), the measured polarisation is:

\begin{equation}\label{eq:chitrue}
\chi_{gal}^{true}=(1+1/R)\left( \chi^{obs}_{gal} - \frac{\chi_{psf}^{true}}{1+R}\right)\;.
\end{equation}

\noindent{However, the contamination from galaxies biases the PSF model, and the measured galaxy polarisation is rather}

\begin{equation}\label{eq:chitbias}
\chi_{gal}^{measured}=(1+1/R')\left( \chi^{obs}_{gal} - \frac{\chi_{psf}^{biased}}{1+R'}\right)\;,
\end{equation}
where $\chi_{psf}^{biased}$ is given by equation~\ref{eq:chibiasx}.

As a result, the measured polarisation ca be written as
\begin{equation}
\chi^{measured}_{gal}=(1+m)\chi_{gal}^{true}+c\;,
\end{equation}
where
\begin{equation}
m=\frac{R/R'-1}{R+1}\;,
\end{equation}
and
\begin{equation}
c=\left(1+\frac{1}{R'}\right)\left(\frac{\chi^{true}_{psf}}{1+R}-\frac{\chi^{biased}_{psf}}{1+R'}\right)
\end{equation}

\noindent{The same relation can be written for the shear:}
\begin{equation}
\gamma^{measured}_{gal}=(1+m')\gamma_{gal}^{true}+c'\;,
\end{equation}
where $m'=m$ and $c'=2c$.

The SExtractor parameters $A_{image}$ and $B_{image}$ can be used to estimate the typical polarisations. In particular, $\chi^{biased}_{psf}$ can be computed as 
\begin{equation}\label{eq:chibias}
\chi_{psf}^{biased}(f_g)=f_g\chi^{mis,gal}+(1-f_g)\chi_{psf}^{true}\;,
\end{equation}
where 
\begin{itemize}
\item $\chi_{psf}^{true}\approx2\gamma_{psf}$ and $\gamma_{psf}$ is estimated as the ellipticity of the PSF, 
\item $\chi^{mis,gal}=\frac{A_{image}^2-B_{image}^2}{A_{image}^2+B_{image}^2}$, for the misclassified galaxies.
\end{itemize}

$R$ and $R'$ can be computed using the SExtractor parameter $Flux\_Radius$:
\begin{equation}
R=\frac{\overline{Flux\_Radius^{gal.}}}{\overline{Flux\_Radius^{stars}}}
\end{equation}
where $Flux\_Radius^{stars}$ is the $Flux\_Radius$ parameters for the true stars,  and 
\begin{equation}
R'=\frac{\overline{Flux\_Radius^{gal.}}}{\overline{Flux\_Radius^{stars+gal_{mis}}}}
\end{equation}
where $Flux\_Radius^{stars+gal_{mis}}$ is the $Flux\_Radius$ parameter for all the objects in the sample labelled as stars (i.e. true stars and misclassified galaxies). 
Using equations~\ref{eq:chitrue} and~\ref{eq:chibias}, we can compute $\chi^{measured}_{gal}$ and $\chi_{gal}^{true}$, as well as the multiplicative and additive biases, $m$ and $c$. Previous work by the DES collaboration led to the formulation of requirements on the value of $m$ and $c$:  $m<0.004$ and $|c|<8\cdot 10^{-4}$. These requirements translate into requirements on the contamination from galaxies. In particular, within a toy model in which $m$ and $c$ depend linearly on $f_g$, i.e. $m=A_mf_g+B_m$ and $c=A_cf_g+B_c$, the expected\footnote{I.Sevilla's personal communication.} values of the parameters are given by $A_m=8.6\cdot 10^{-2}$, $B_m=-1.6\cdot10^{-3}$, $A_c=-1.0\cdot 10^{-1}$ and $B_c=2.1\cdot10^{-3}$.
Therefore, the requirement $m<0.004$ translates into $f_g<f_{g,lim}$ with $f_{g,lim}=0.07$ i.e. $p^s>p^s_{lim}$ with $p^s_{lim} =93\%$. The requirement on the additive bias parameter $|c|<8\cdot 10^{-4}$ leads to a more stringent requirement on the contamination: $f_{g,lim}=0.03$, i.e. $p^s_{lim}=97\%$.

In practice, shear codes have the ability to sharpen the classification of stars and galaxies. Indeed, a shear measurement code convolves a model for the galaxy with the measured PSF function, and then adjusts the parameters of this model to best fit the observed data. If, for example, the best-fit values for the parameters characterising the size of the model are too small, it is likely that the observed object is a star (or a very small galaxy). This allows to perform additional cuts of the sample of objects, using the output of the shear measurement code as an additional indication about the class of the object. For this reason, using the derived verbatim as a requirement on the star/galaxy separation is conservative.

\subsection{Summary of the science requirements star/galaxy separation}
The requirements on the quality of the star/galaxy separation derived in this section are summarised in table~\ref{req}.

\begin{table}\caption{Summary of the science requirements on the quality of star/galaxy separation.} \label{req}
\centering
 \begin{tabular}{|| l || l || l ||}
  \hline
\rule[-2.ex]{0pt}{5ex}&{\bf LSS} & {\bf WL} \\
\hline
 \rule[-2.ex]{0pt}{5ex}$p^g$ & $\ge 98.5\%$  (requirement & $\ge p^g_{lim}$, with \\
\rule[-2.ex]{0pt}{5ex}& on the systematic error)& $p^g_{lim}\in[87.7\%,97.8\%]$\\
\rule[-2.ex]{0pt}{5ex}&& (req. on the systematic error) \\
 \hline
\rule[-2.ex]{0pt}{5ex}$p^s$&-&$>97\%$ \\
\rule[-2.ex]{0pt}{5ex}&&(req. on the PSF calibration)\\
 \hline
\rule[-2.ex]{0pt}{5ex} $c^g$& $\ge 88.9\%$ (requirement& $\ge 96.0\%$ (requirement\\
\rule[-2.ex]{0pt}{5ex} &on the statistical error)&on the statistical error)\\
 \hline
\rule[-2.ex]{0pt}{5ex} $c^s$& - &$\ge 25\%$ (requirement\\
\rule[-2.ex]{0pt}{5ex}&&on the PSF calibration)\\ 
\hline
\end{tabular}
\end{table}

A dedicated sample of stars is only needed when calibrating the PSF. Therefore, the two requirements on the samples of stars are only required for WL science. As far as samples of galaxies are concerned, LSS science requires purer samples than WL science. This is due to star contamination affecting the corresponding measured ``observable'' in different ways. The contribution of misclassified stars to the measured shear is dominated by the shot noise term (see Eq.~\ref{eq:tile}), which is approximately scale independent, whereas they mimic a l-dependent density fluctuation of galaxies and therefore contribute to the LSS measurement in a more complicated way. On the other hand, WL requires a more complete samples of galaxies. This is because a ``usable'' object means something different for LSS and WL. In order to be usable for LSS measurement, a galaxy needs to be detected with a reliable photometric redshift but WL also needs the shape of the galaxy to be measurable.

In the next sections, we will use these requirements to assess the performance of a new classifier, multi\_class, and compare it to other classifiers currently used in galaxy surveys. In particular, we will use the most stringent requirement, in the cases of the purities $p^g$ and $p^s$, i.e. $p^g_{lim}=97.8\%$ and $p^s_{lim}=97\%$. 

\section[]{Current tools for star-galaxy\\* Separation}

Different strategies have been adopted to classify stars and galaxies in large sky surveys. The morphometric approach (e.g. \citealt{Kron}; \citealt{Yee}; \citealt{Vasconcellos}; \citealt{Sebok}, \citealt{Valdes}) relies on the separation of point sources (the ones most likely to be stars) from resolved sources (presumably galaxies).

This approach is challenged at the faint magnitudes reached by the next generation of wide-field surveys, due to the vast number of unresolved galaxies. 

Another strategy consists of using training algorithms. Machine learning distinguishes several types of learning strategies, Artificial Neural Network (ANN) being one successfully implemented example of {\it supervised learning}. ANN has previously been applied to the star/galaxy separation problem (e.g. \citealt{odewahn}, \citealt{Naim},  \citealt{ba}, \citealt{Oyaizu}). Indeed, star/galaxy separation shares with many other classification problems the three criteria which usually make neural computing applications particularly successful: 
\begin{itemize}
\item The task is well-defined in that we know precisely what we want, i.e. classify objects in two distinct classes.
\item There is a sufficient amount of data available to train the network to acquire a useful function based on what it should have done in these past examples.
\item No simple parametrization for the output (the class of the object) as a function of the input (the parameters derived from the images) is known, and we would like to leave it to the algorythm to determine the optimal classification scheme.
\end{itemize}
Other supervised classifiers, such as Support Vectors Machine (SVM), have been more recently used for the star/galaxy separation problem, as well as unsupervised tools such as Hierarchical Bayesian techniques (e.g. \citealt{Fadely,Henrion}).

Throughout this section, we will use the following notations to define:
\begin{itemize}
\item  {\it classification tools} - class\_star; spread\_model and multi\_class
\item {\it classification output} - $X_{class\_star}$; $X_{spread\_model}$ and $X_{multi\_class}$ .
\end{itemize}
As described below (section $4.1$),  class\_star and spread\_model are two classifiers currently implemented in {\it SExtractor} \citep{ba} and in the next sections we present a new method for star/galaxy separation called ``multi\_class'', designed to achieve the science requirements derived in section $3$ at the faint magnitudes reached by DES. 

\subsection{Current approaches}
Both the morphometric and the training approaches are implemented in {\it SExtractor} (\citealt{ba}), with two classifiers, class\_star and spread\_model.
\begin{figure}
\includegraphics[width=8cm]{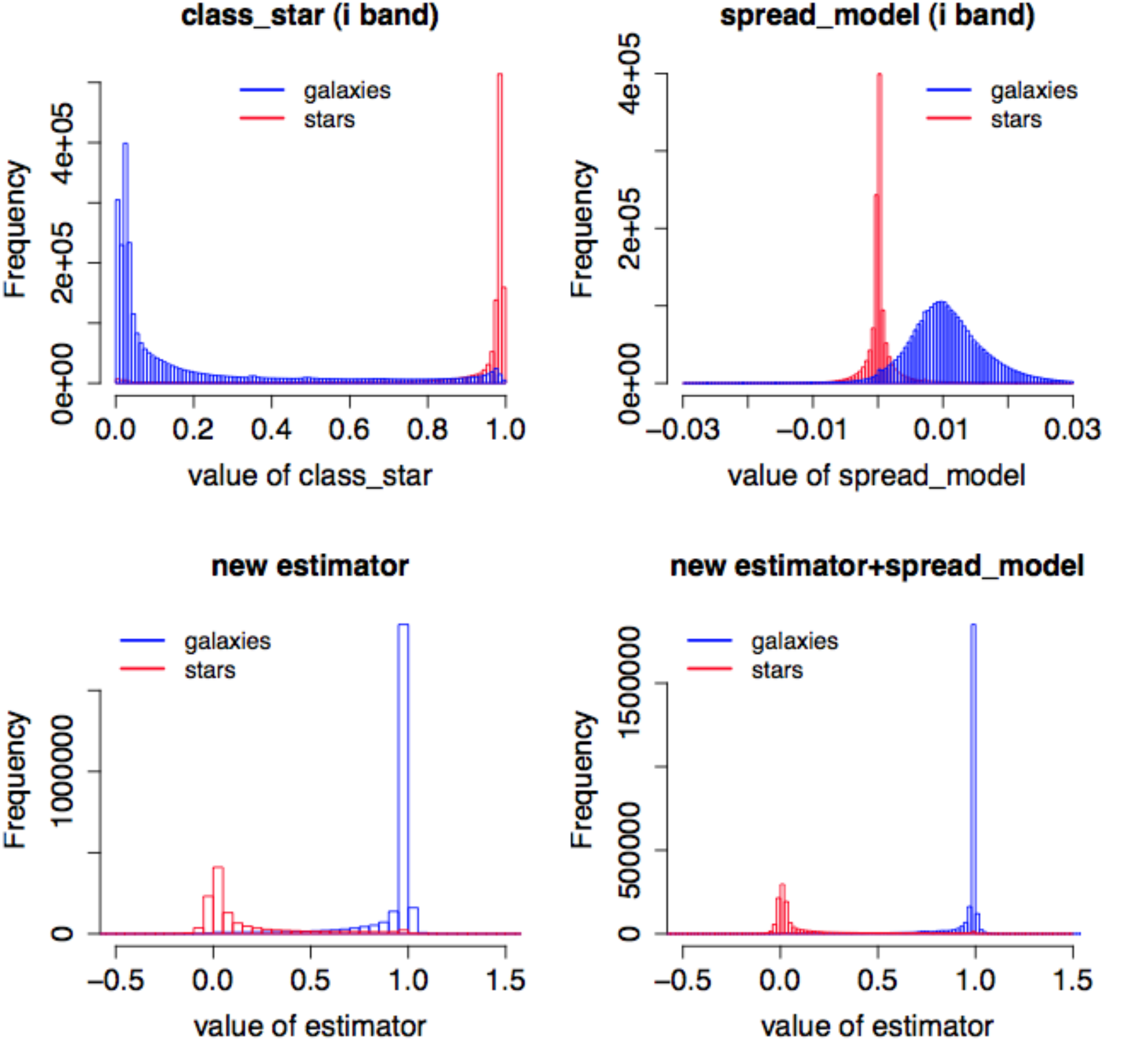}
\caption{Distribution of the output of all the classifiers presented in the paper. The two upper histograms show the classification performed by class\_star and spread\_model. The lower histograms show the classification performed by our new estimator, multi\_class. On the right one, we incorporate $X_{spread\_model}$ in the input parameters of the ANN. The advantages of plugging $X_{spread\_model}$ into our tool are explained in section $5.3.2$. This allows an increase of the purity for a given completeness, as shown in figure~\ref{fig:Emmanuel}.}
\label{fig:histo_all_classifiers}
\end{figure}
\subsubsection{The training approach - class\_star}

The first classifier to be implemented in {\it SExtractor} was class\_star. Its performance on our example sub-survey is shown in figure~\ref{fig:histo_all_classifiers}. It uses a set of features of the objects as the input space for a built-in  previously trained ANN. These parameters are:
\begin{enumerate}
\item eight isophotal areas, at regular intervals spanning from the detection threshold to the intensity peak;
\item the intensity peak;
\item the local value of the seeing.
\end{enumerate}

This specific pre-defined set of inputs, chosen mainly for historical reasons, is the main weakness of the class\_star estimator. The choice of training the ANN on isophotal areas (normalised to the local PSF footprint area) makes it sensitive to close pairs of objects (star-star, star-galaxy, galaxy-galaxy) either blended or de-blended. Since star-star pairs are common on the bright end of the source population, the classifier has a tendency to miss bright, compact galaxies. 

More generally speaking, given the large amount and diversity of information encapsulated in the parameters provided by {\it SExtractor}, this specific choice of inputs has become hard to justify as it is using a very small part of the available information.  The photometry, the shape or the size of an object should also be useful indicators of whether it is a star or a galaxy.

Class\_star has the advantage of making use of several parameters and combining the information they contain. In this sense it is a ``multi-parameter" estimator. However, it does not use the most relevant parameters. A more flexible and sensible choice of the inputs is likely to give much better results. This is the main motivation for the new approach tested in this paper.

\subsubsection{The morphometric approach - spread\_model} 
 The morphometric approach was used in several photometric surveys in the past. One possible implementations of this approach, adopted in the Sloan Digital Sky Survey (SDSS) pipeline and in early versions of the DES pipeline, consists of comparing a ``model magnitude'', i.e. the optimal measure of the magnitude obtained by fitting a galaxy model to the object, to the ``PSF magnitude'' , i.e. the optimal measure of the magnitude determined by fitting a PSF model to the object. A similar strategy was adopted in the Canada-France-Hawaii Telescope Legacy Survey (CFHTLS) pipeline, where classes are assigned to objects according to their half-light radius (HLR), i.e. the circular radius which encloses half the light of an object.

The classifier implemented in recent development versions of {\it SExtractor} is called spread\_model (\citealt{desai}, \citealt{Bertin}). It carries out diverse operations directly on the image pixels with no use of the object's parameters generated by {\it SExtractor}. 
The newest version of spread\_model acts as a linear discriminant between the best fitting local PSF model {\bf $\phi$} and a slightly ``fuzzier" version made from the same PSF model, convolved with a circular exponential model with scale length given by $FWHM/16$ (FWHM being the Full-Width at Half-Maximum of the local PSF model). Spread\_model is normalized to allow for comparison of sources with different PSFs throughout the field. It is defined as

\begin{equation}\label{eq:spread}
\mathrm{X_{spread\_model}}= \frac{\phi^TWx}{\phi^TW\phi} - \frac{G^TWx}{G^TWG}\;,
\end{equation}
where $x$ is the image centred on the source, $W$ is the inverse of the covariance matrix of the pixel noise, which is assumed to be diagonal, {\bf $\phi$} is the PSF and G is the circular exponential model convolved with the PSF. By construction, spread\_model is close to zero for point sources (most likely to be stars), positive for extended sources (most likely to be galaxies) and negative for detections smaller than the PSF, such as cosmic ray hits. 

The performance of this late version of spread\_model on our example sub-survey is shown in figure~\ref{fig:histo_all_classifiers}. Although this morphometric approach is quite efficient, it is not entirely satisfying as it does not make use of any of the $300$ {\it SExtractor} parameters, which are likely to encapsulate a lot of relevant information for star/galaxy separation. 

\section[]{The multi\_class method}
\subsection{Motivation and principle}

Our goal is to combine the assets of both the morphometric approach and the training approach. We adopt the multi-parameter approach allowed by the training method and focus on making the optimal choice of input parameters.
The steps of the method are as follows:

\begin{enumerate} 
\renewcommand{\theenumi}{(\arabic{enumi})}
  \item  Optimal choice of input parameters using a PCA;
  \item  Training and running an ANN.
  \end{enumerate} 

\subsection{Step 1- optimal choice of input parameters using Principal Component Analysis }

We make a broad pre-selection of all the parameters likely to be relevant for star/galaxy classification. These parameters are listed in table~\ref{presel}. They include:
\begin{enumerate}
\item {\bf photometry} in 5 bands ({\it g},{\it r},{\it i},{\it z} and {\it y});
\item  the {\bf size} of objects;
 \item the {\bf shape} of objects;
 \item the {\bf surface brightness} of objects;
 \item qualifiers of the {\bf fitting} procedure;
 \item the {\bf output of the class\_star classifier}, $X_{class\_star}$;
 \item additional {\bf analysis-dependent information}.
 \end{enumerate}
\begin{table}
 \caption{DC6 pre-selected parameters, grouped as defined in section $5.2$, by type of information they provide: (i): photometry; (ii) size; (iii): shape; (iv): surface brightness; (v): qualifiers of the fitting procedure; (vi): output of the class\_star classifier; (vii): additional analysis-dependent information. It should be noted that all of these parameters are distance-dependent. The need for K-correction to the magnitudes is therefore dealt with by including the photometric redshift in this pre-selected parameters space.}
 \label{presel}
  \begin{tabular}{|| l || l || l ||}
  \hline
&{\bf Parameters} & {\bf Description} \\
 \hline
(i)&mag\_aper\_ in 5 bands & Fixed aperture magni- \\
&& -tude with 6 \\
&&different apertures \\
&mag\_auto in 5 bands & Kron-like elliptical \\
 && aperture magnitude \\
&mag\_iso in 5 bands & Isophotal magnitude \\
&mag\_model in 5 bands & Magnitude from model-\\
&&fitting \\
&mag\_petro in 5 bands & Petrosian-like elliptical\\
&&aperture magnitude \\
&mag\_psf in 5 bands & Magnitude from PSF-\\
 &&fitting  \\
&mag\_spheroid in 5 bands & Spheroid total magn-\\
&&-itude\\
 && from fitting \\
\hline
(ii)&kron\_radius (from the de-& Kron apertures\\
&tection image)&\\
\hline
(iii)&ellipticity (from the de-& $1- B_{image}/A_{image}$\\
&tection image)&\\
\hline
(iv)&isoarea\_world in 5 bands & Isophotal area above \\ 
&&analysis threshold\\
&FWHM\_world in 5 bands &FWHM assuming a \\
&&gaussian core\\
\hline
(v)&chi2\_model in 5 bands &   Reduced chi-square \\
&&of the fit\\
&chi2\_psf in 5 bands & Reduced chi-square from \\
 &&PSF-fitting\\
&niter\_model in 5 bands & Number of iterations for \\
&&model-fitting \\
\hline
(vi)&$X_{class\_star}$ in 5 bands & Output from\\
 && class\_star \\
 \hline
(vii)&nlowdweight\_iso & Number of pixels with low\\
&& detection weight over the\\
&& isophotal profile\\
&photoZ & photometric redshift \\ 
 \hline
 \end{tabular}
\end{table}

Ideally, we could run an ANN with this full set of relevant inputs. In practice, training the ANN is a non-linear iterative process, which becomes more time consuming and less robust as the number of input parameters increases. In fact, defining an optimal set of input parameters consists of minimising its size while maximising the amount of relevant information it contains. 

\begin{figure}
\includegraphics[width=8cm]{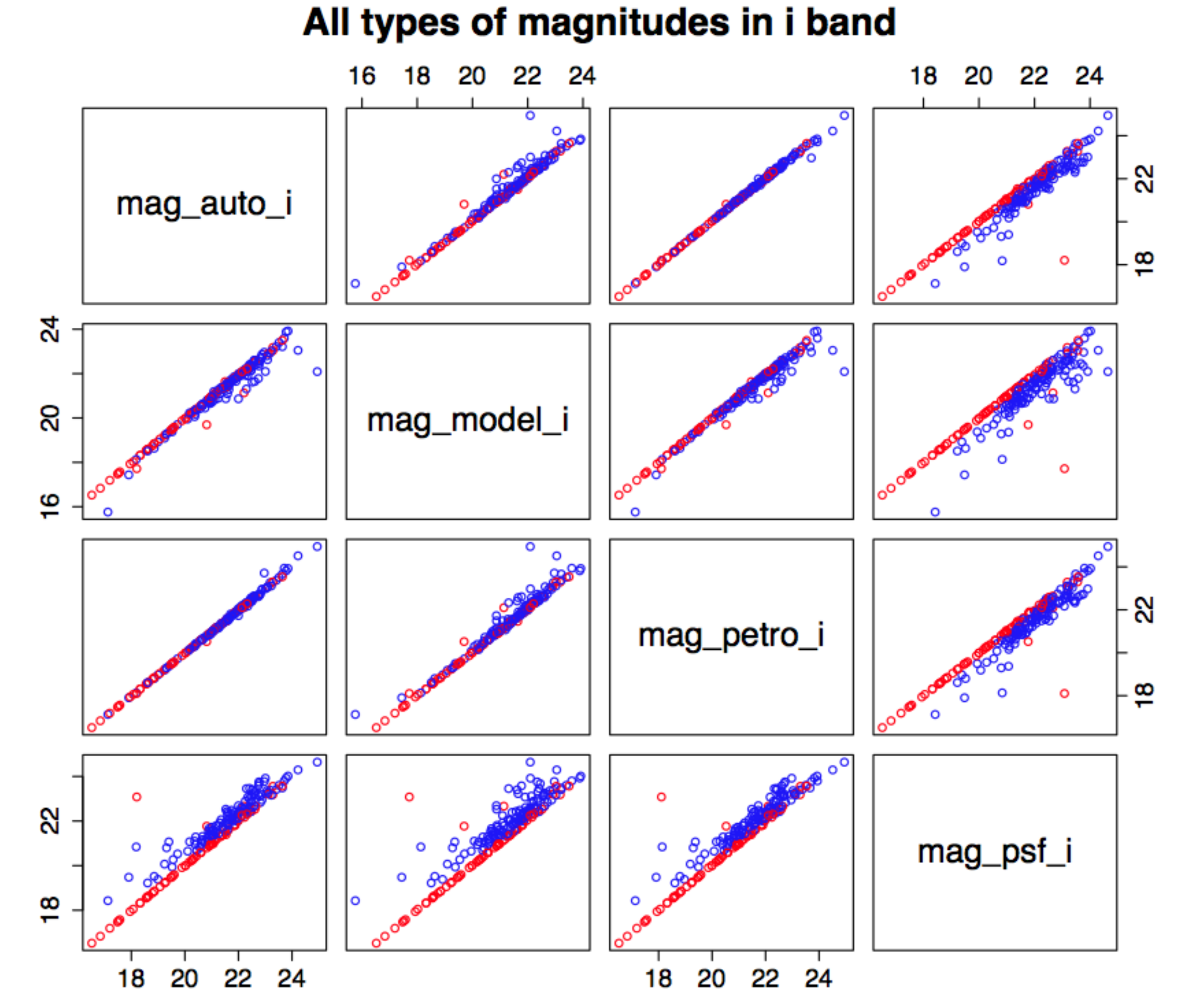}
\caption{Scatter plots for stars (red markers) and galaxies (blue markers), for four different types of magnitudes in the {\it i} band. The magnitudes are strongly correlated and PCA is therefore well adapted to re-express them in a new basis of independent variables.}
\label{fig:scatter}
\end{figure}

Our initial set of parameter is redundant, as many of the parameters within each sub-group are dependent variables. For example, we show in figure~\ref{fig:scatter} the dependencies between four types of magnitudes parameters measured in a given band. In order to reveal the redundancies within the data and compress it, we use a Principal Component Analysis (PCA). This statistical method, which comes down to diagonalising the covariance matrix of the data, allows us to re-express the pre-selected parameters detailed above in a more meaningful basis of orthogonal, i.e. uncorrelated variables called {\it principal components}. The first principal component is chosen to account for most of the data variability and thus to have the highest possible variance. Then each succeeding principal component has the highest possible variance under the constraint of being orthogonal - that is uncorrelated - to the preceding one.

We run several ``well-informed'' PCAs on sub-ensembles of parameters, rather than a ``blind'' PCA on the full set of initial parameters. We choose to group in these sub-ensembles parameters which have the same units (or measure) and which are linearly dependent on each other (such as the magnitudes in a given band, as shown in figure~\ref{fig:scatter}). Indeed, when the parameters are linearly dependent, PCA is successful at finding a new basis of meaningful independent variables.

Our new set of parameters includes uni-band parameters from the initial set (such as the photometric redshift or the ellipticity), as well as the principal components from the PCAs listed below:
\begin{itemize}
\item PCA on the five bands of each multi-band parameter;
\item PCA on the six fixed-aperture magnitudes in each band;
\item PCA on the six other types of magnitudes in each band (i.e. {\it mag\_auto}, {\it mag\_iso}, {\it mag\_model}, {\it mag\_petro}, {\it mag\_spheroid} and {\it mag\_psf}).
\end{itemize}

\begin{figure}
\includegraphics[width=8cm]{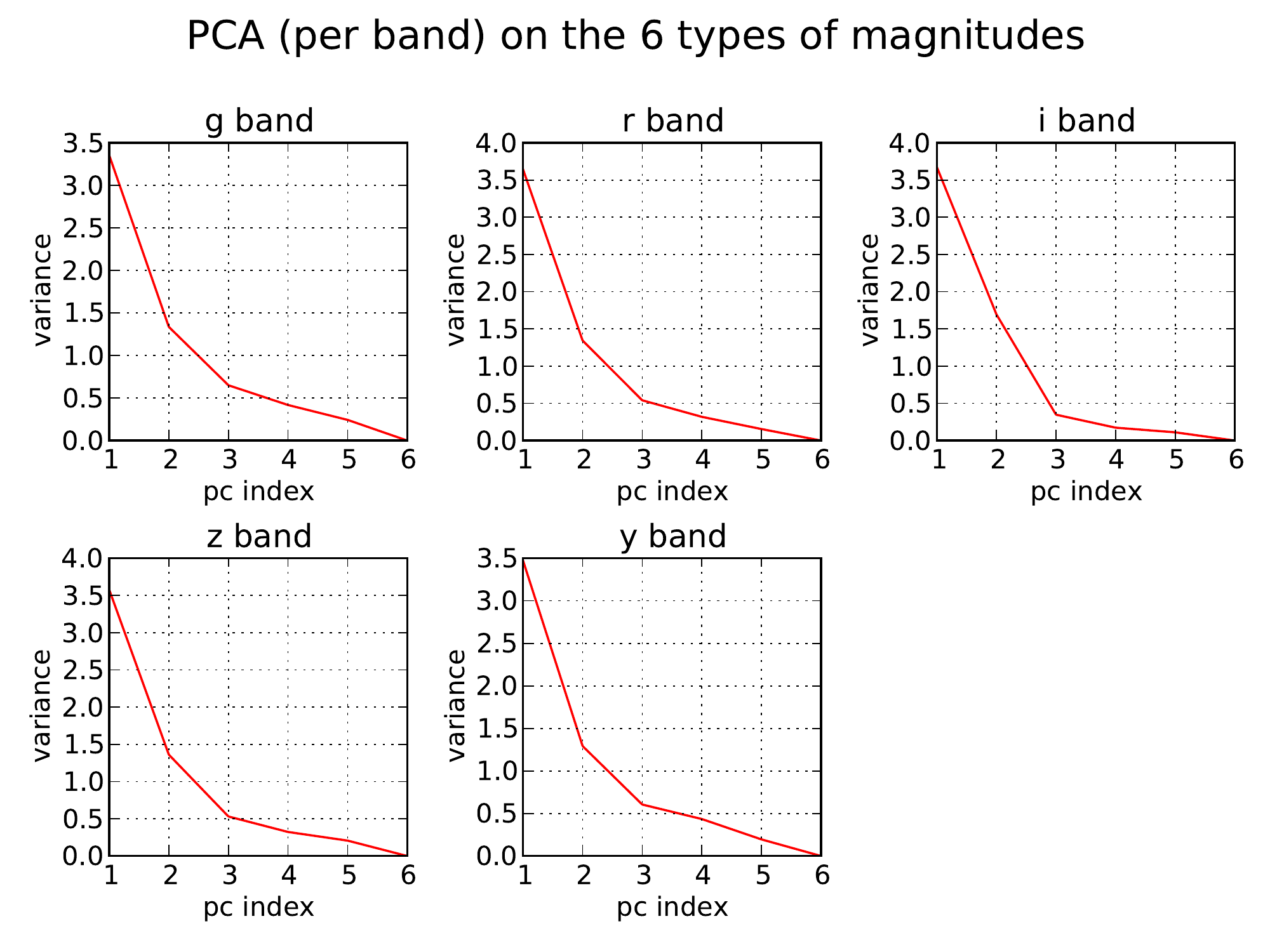}
\caption{Value of the variance of the principal components as a function of their index for the fives (per-band) PCAs performed on the six types of magnitudes: {\it mag\_auto}, {\it mag\_iso}, {\it mag\_model}, {\it mag\_petro}, {\it mag\_spheroid} and {\it mag\_psf}.}
\label{fig:variance}
\end{figure}

\begin{center}
\begin{figure*}
\includegraphics[width=8cm]{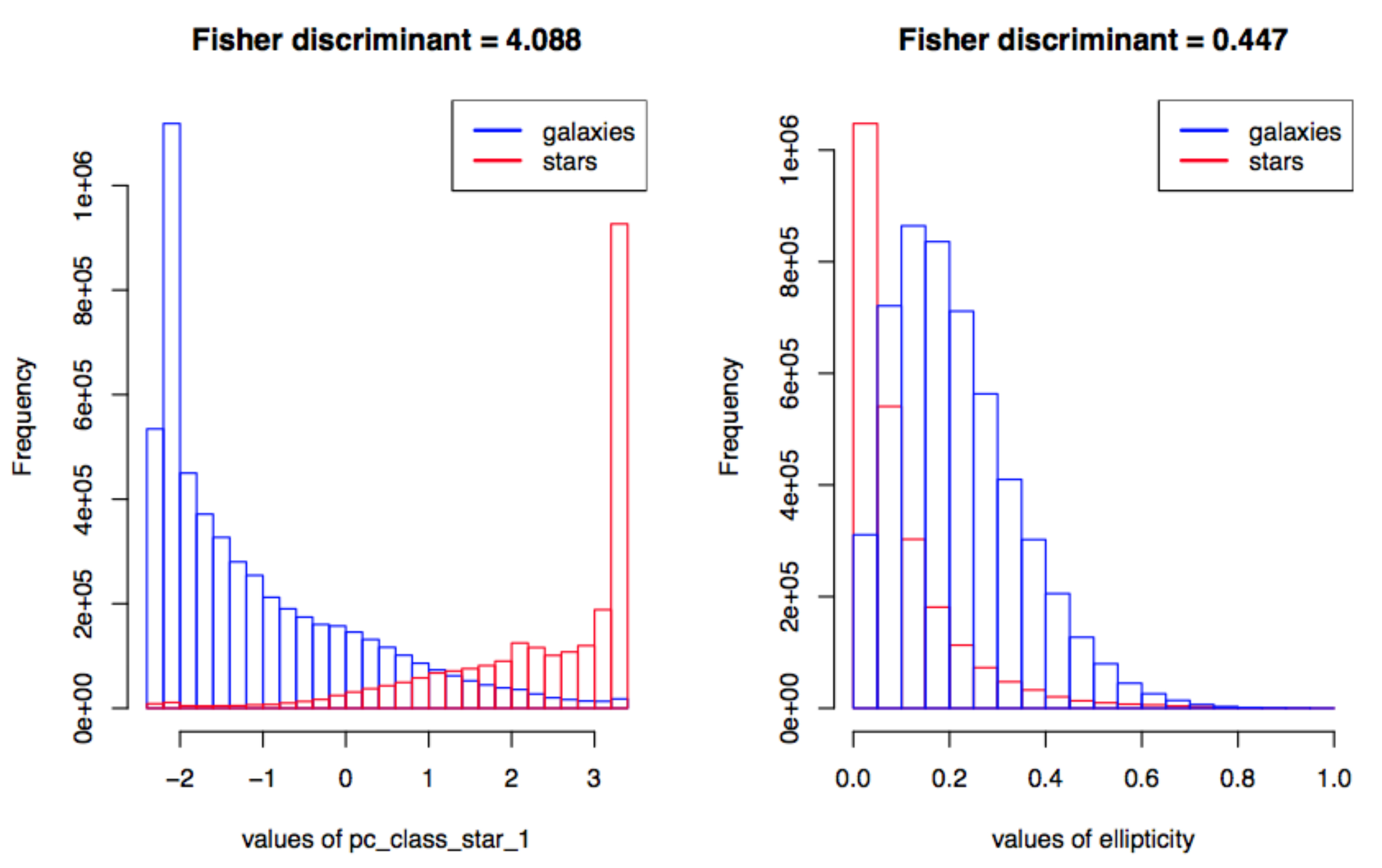}
\includegraphics[width=4.2cm]{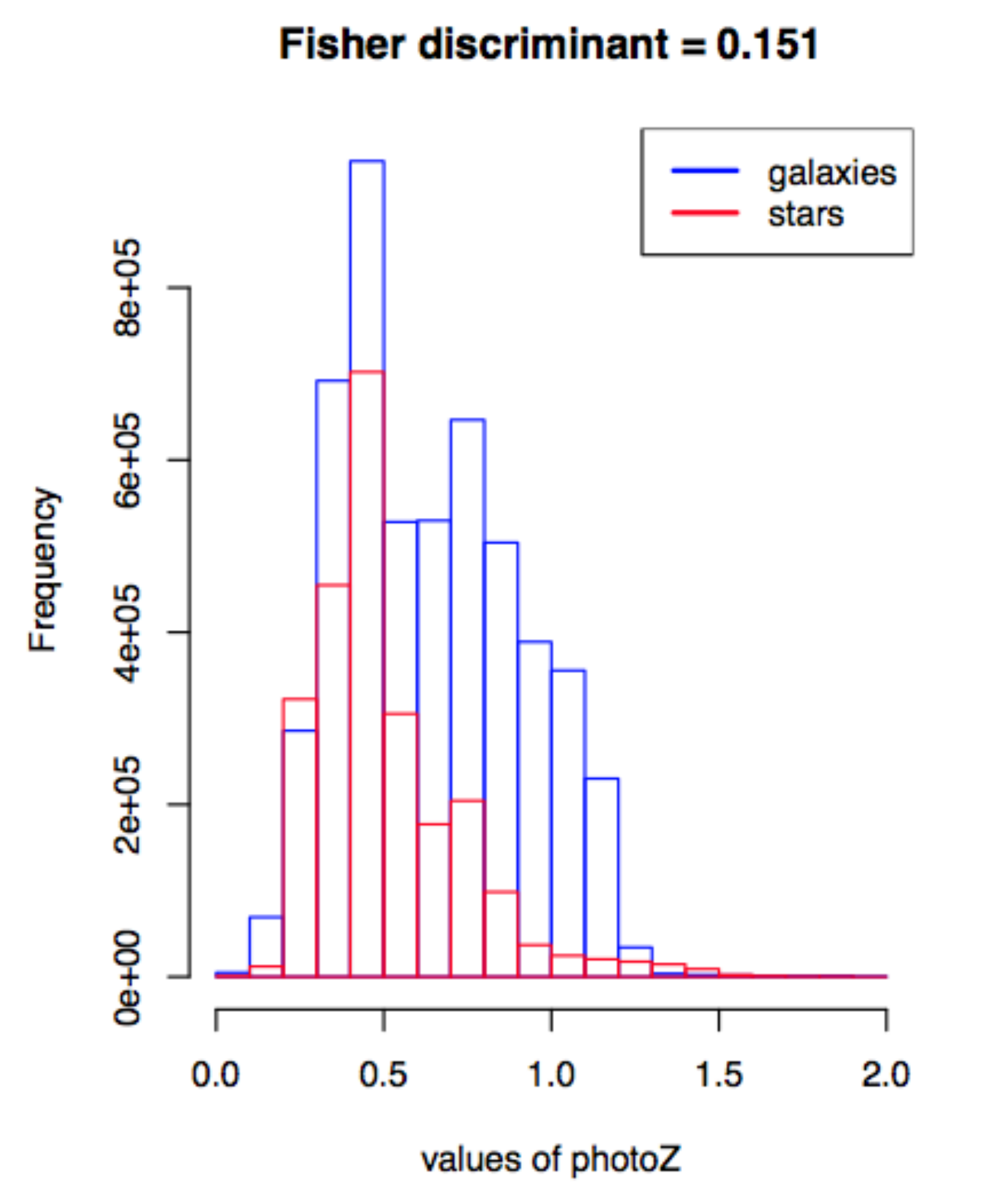}
\caption{Distribution of the three parameters with the highest Fisher discriminant, for stars and galaxies as indicated in the figure. {\it pc\_class\_star\_1} (left) is the first principal component from a PCA performed on the five bands of $X_{class\_star}$ (see section $5.3.2$).  The two other parameters shown, {\it ellipticity} (centre) and {\it photoZ} (left) have not gone through any PCA.}
\label{fig:best_fish}
\end{figure*}
\end{center}

Figure~\ref{fig:variance} shows the variances of the principal components of these six types of magnitudes in each band as a function of their index. Each of these PCAs shows that most of the variance of the data is encapsulated in a reduced number of principal components. In many cases, using PCA for data reduction consists of selecting only the principal components with the highest variance and approximating the data by its projection on this smaller set of variables. This encompasses the assumption that the important information is represented by the components with the highest variances. In the case of star/galaxy separation, this assumption is too simplistic. Indeed, the class of an object is only one possible source of variance and high variance could also be due to differences between objects in a given class. 
Therefore, when looking for the most relevant components for star/galaxy separation, we need another criterion to quantify their aptitude to separate between the classes. We calculate the {\it Fisher discriminant} \citep{Fish} for each of the new parameters, defined as the inter-class variance over the intra-class variance,
\begin{equation}\label{eq:Fisher}
\mathcal{F}_i=\frac{(\overline{X_{G,i}}-\overline{X_{S,i}})^2}{\sigma_{G,i}^2+\sigma_{S,i}^2}\; , 
\end{equation}
where $\overline{X_{A,i}}$ is the empirical mean value of the $i^{th}$ parameter for class A and $\sigma^2_{A,i}$ is its empirical variance. 
Figure~\ref{fig:best_fish} shows the classification performed by the three parameters with the highest Fisher discriminant.
The fifteen parameters with the highest Fisher discriminant form our final set of input parameters for the ANN (as dicussed in section $5.3.1$, more than twenty input parameters make the ANN less robust, so we limit the basic set to fiftenn parameters, in anticipation of the other five that will be added in section $5.3.2$).

The set of parameters with the highest Fisher discriminant is data-specific. In our analysis of DC6, the most discriminative combinations of parameters are those of the class\_star different bands, (group (vi) of table~\ref{presel}), morphometric parameters such as the ellipticity (group (ii) of table~\ref{presel}), followed by photometric parameters such as the photometric redshift and the magnitudes (group (i) of table~\ref{presel}). This being said, when generalising the method to other data sets, the Fisher discriminant should be recalculated.
\subsection{Step 2 - running an ANN on the optimal inputs space}

Once a set of optimal parameters is defined, the next step consists of mapping these parameters to the class of the objects. This mapping is performed by training an ANN.

\subsubsection{ANN: principle and advantages}

In essence, an ANN is a highly-flexible, fully non-linear fitting algorithm. 
During the training phase, it receives a set of input patterns and a given property (in our case the class of the object), which needs to be fitted to them. The training consists of several iterations during which a number of free parameters known as {\it weights} are adjusted so as to minimise the difference between the outputs of the neural network for each pattern and the desired property. The algorithm then learns how to link the inputs to the desired property. After the training phase, the ANN can be used to infer this property from a set of input objects for which it is unknown.
For our analysis, we train an ANN to map the set of optimal input parameters selected in section $5.2$ to the class of the object (star or galaxy) on a sample of objects for which the answer is known (the training is made on the DC6 simulations for which we know the true class of each object). The ANN is then used to deduce the class of a distinct set of objects.

An ANN is made of computing units called {\it neurons}, arranged in several layers and connected by synapses in which the information flows in a single direction. The complexity of the network depends on the number of layers and neurons in each layer. We chose to use the ANNz photometric redshift code \citep{cl4} , which was originally designed for photometric redshift measurements, but can be effectively and straightforwardly applied to our classification problem. The trade-off between the complexity of the network and its performance has been investigated by \citet{f}. For the same number of parameters, adding extra hidden layers is found to give greater gains than widening existing layers. As the network complexity is increased, the accuracy eventually converges so that no further improvement is gained by adding additional nodes. We chose a network architecture with an input layer of fifteen parameters (or twenty, as explained in the next section) and two hidden layers of twenty nodes, which turns out to be sufficiently complex for such convergence to be achieved. 

Training on real data, as opposed to simulations, is preferable, yet more challenging. One option would be to use data from space-based surveys, as in space the PSF is not affected by the seeing. Data from the Hubble Space Telescope could be used to train our tool for the real DES survey data.

\subsubsection{Plugging other classifiers in the method}
\begin{center}
\begin{figure*}
\includegraphics[width=8cm]{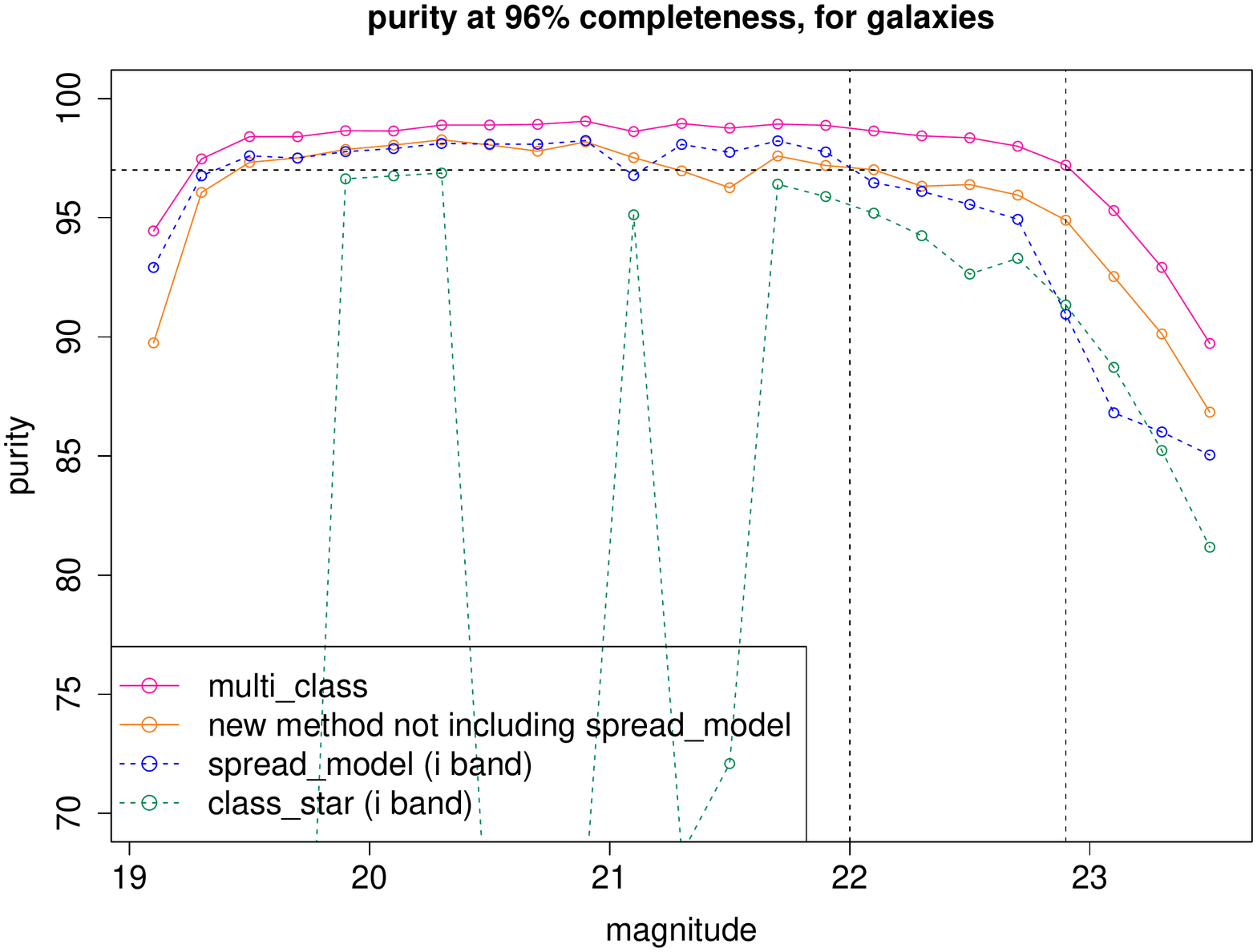}
\includegraphics[width=8cm]{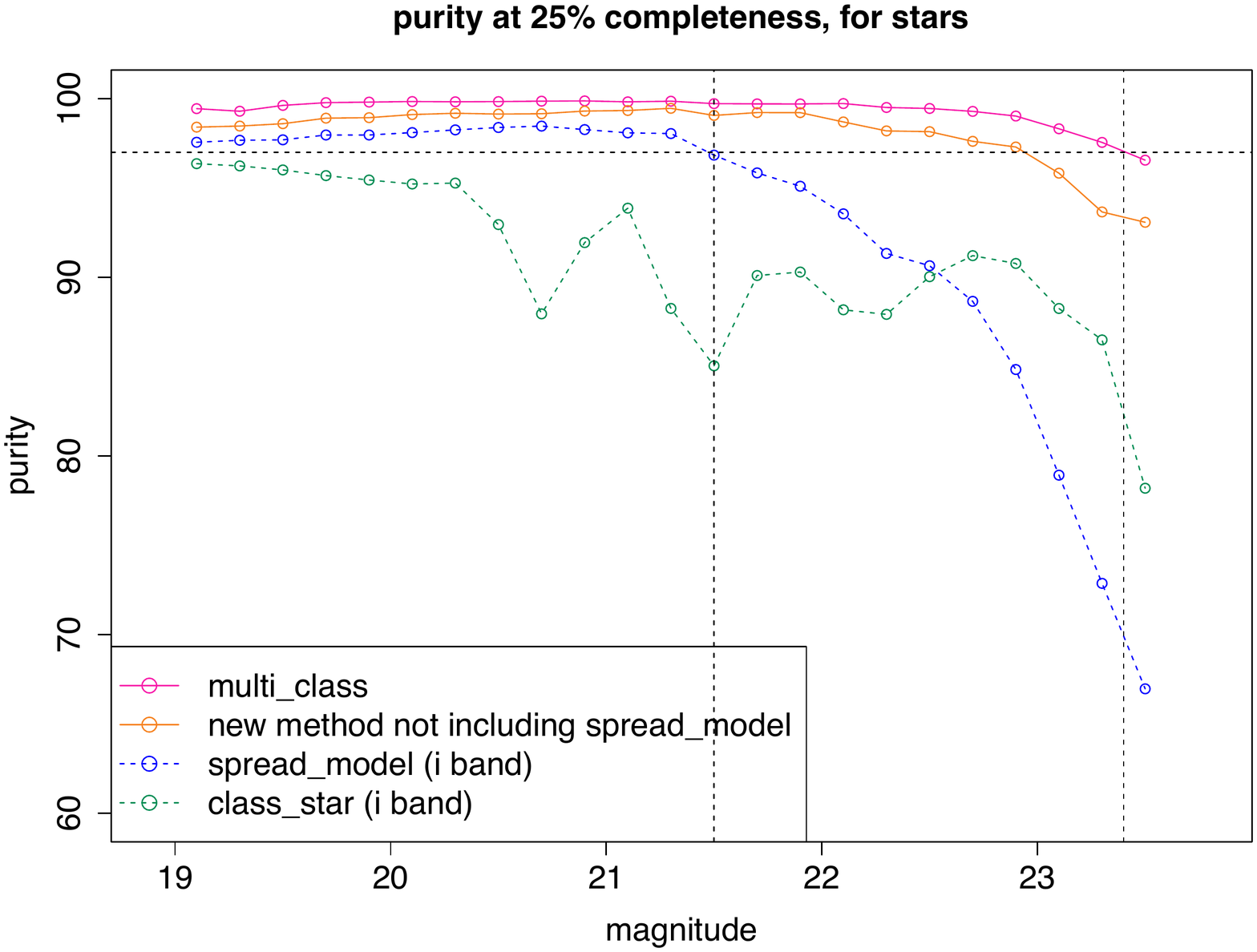}
\caption{Purity level at the required completeness, for the WL probe, as a function of magnitude in the {\it i} band. The orange and pink curves correspond to different versions of our method, whereas the blue and green ones show the performance of the classifiers class\_star and spread\_model. The orange curve is obtained when running the ANN on the 15 parameters selected in section $5.2$ and the pink curve, the final version of multi\_class, is obtained when adding spread\_model in five bands to this set of inputs. The dashed horizontal line shows the science requirement from WL science on $p^g$ ($97.8 \%$, section $3.3.1$) and $p^s$ ($97.0 \%$, section $3.4$). The requirement on $p^g$ is achieved by multi\_class up to magnitudes of $22.9$, whereas spread\_model only allows us to reach $22.0$. The requirement on $p^s$ is achieved up to magnitudes of $23.4$ with multi\_class, versus $21.5$ with spread\_model.}
\label{fig:Emmanuel}
\end{figure*}
\end{center}
Using an ANN brings flexibility to the training approach. It allows us not only to choose which inputs to use, but also in what number. In particular, we can take the output of other classifiers as inputs to our method. 

We run a PCA on the five $X_{class\_stars}$ (in the five bands). Not surprisingly, the first principal component has a high Fisher discriminant (as shown in figure~\ref{fig:best_fish}) and is therefore included in the 15 input parameters selected in section $5.2$. As the the five bands of $X_{spread\_model}$ are less clearly linearly dependent, we choose not to run a PCA on them and add the five $X_{spread\_model}$ to the set of fifteen input parameters, which amounts to twenty input parameters. 

Figure~\ref{fig:Emmanuel} presents the purity level at a given completeness for these two different configurations of our method. The performance of our method with fifteen input parameters (orange curve) can be compared to the performance when plugging in $X_{spread\_model}$ (pink curve). Including $X_{spread\_model}$ in the inputs allows an increase in the level of the purity by $2\%$ at faint magnitudes. Running the ANN on the fifteen preselected parameters (orange curve) already gives better results than spread\_model (blue curve) for most of the magnitude range (except for the very faint magnitudes, in the galaxies case).  However, the best results are obtained by combining the two, i.e by running the ANN on a hybrid input space combining the 15 selected parameters and $X_{spread\_model}$.

\section[]{Classification results}

We showed that we can optimise our classifier performance by using a ``well-informed'' PCA strategy (section $5.2$), and by incorporating $X_{spread\_model}$ into the method (section $5.3.2$). We now compare our classifier performance to the one of the other classifiers. We will focus on comparing multi\_class to spread\_model, as the performance of class\_star is widely surpassed by both spread\_model and multi\_class for most of the magnitude range (as shown in Figure~\ref{fig:Emmanuel}).

\begin{figure}
\includegraphics[width=7.8cm]{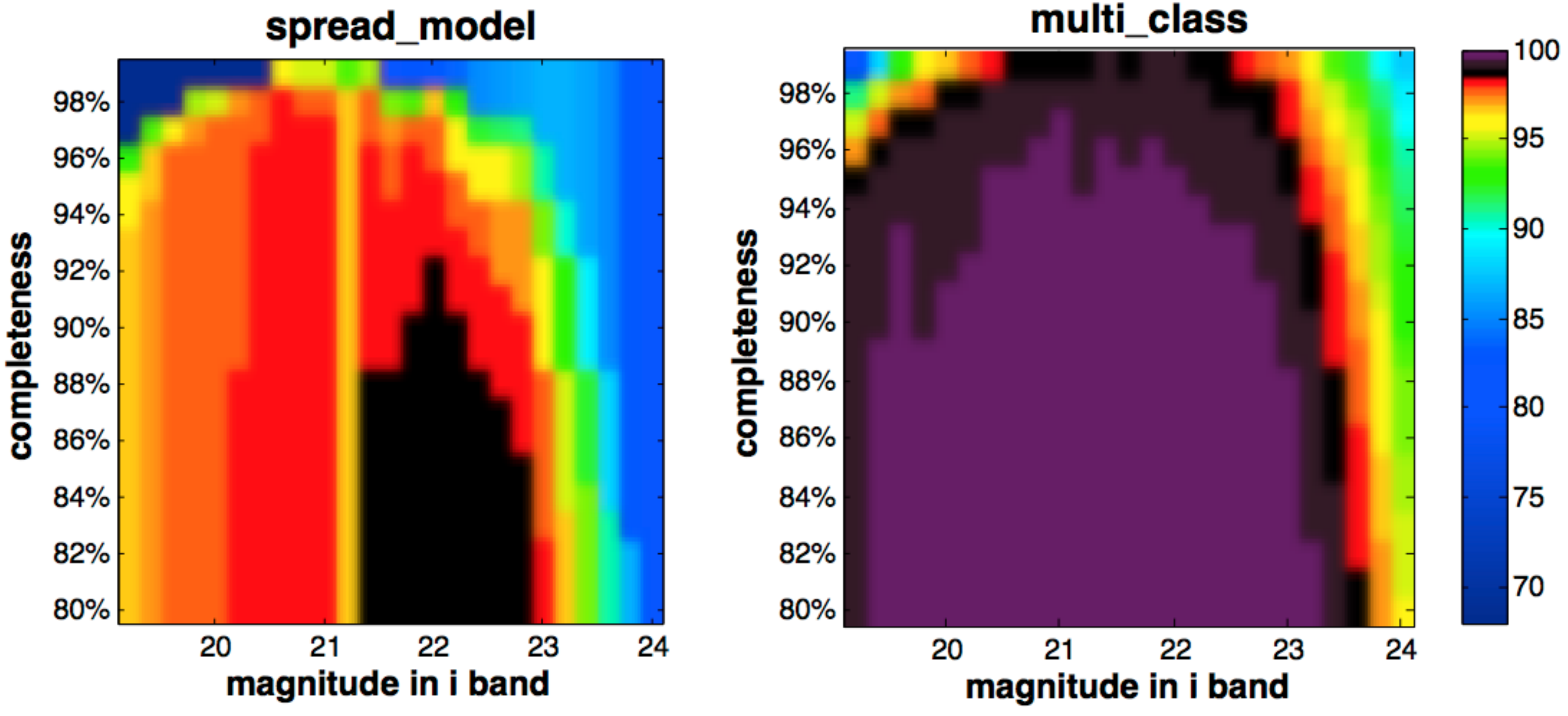}
\caption{Level of purity for a sample of galaxies $p^g$, for different magnitudes and values of the completeness. The $98.5\%$ level requirement from LSS (section $3.3.2$) is shown in purple, and the $97.8\%$ limit required for WL (section $3.3.1$) is shown in black. Spread\_model does not allow to achieve the LSS requirement, which multi\_class can reach. Multi\_class also allows us to achieve the requirement from WL at fainter magnitudes than spread\_model.} 
\label{fig:black_gal}
\includegraphics[width=7.8cm]{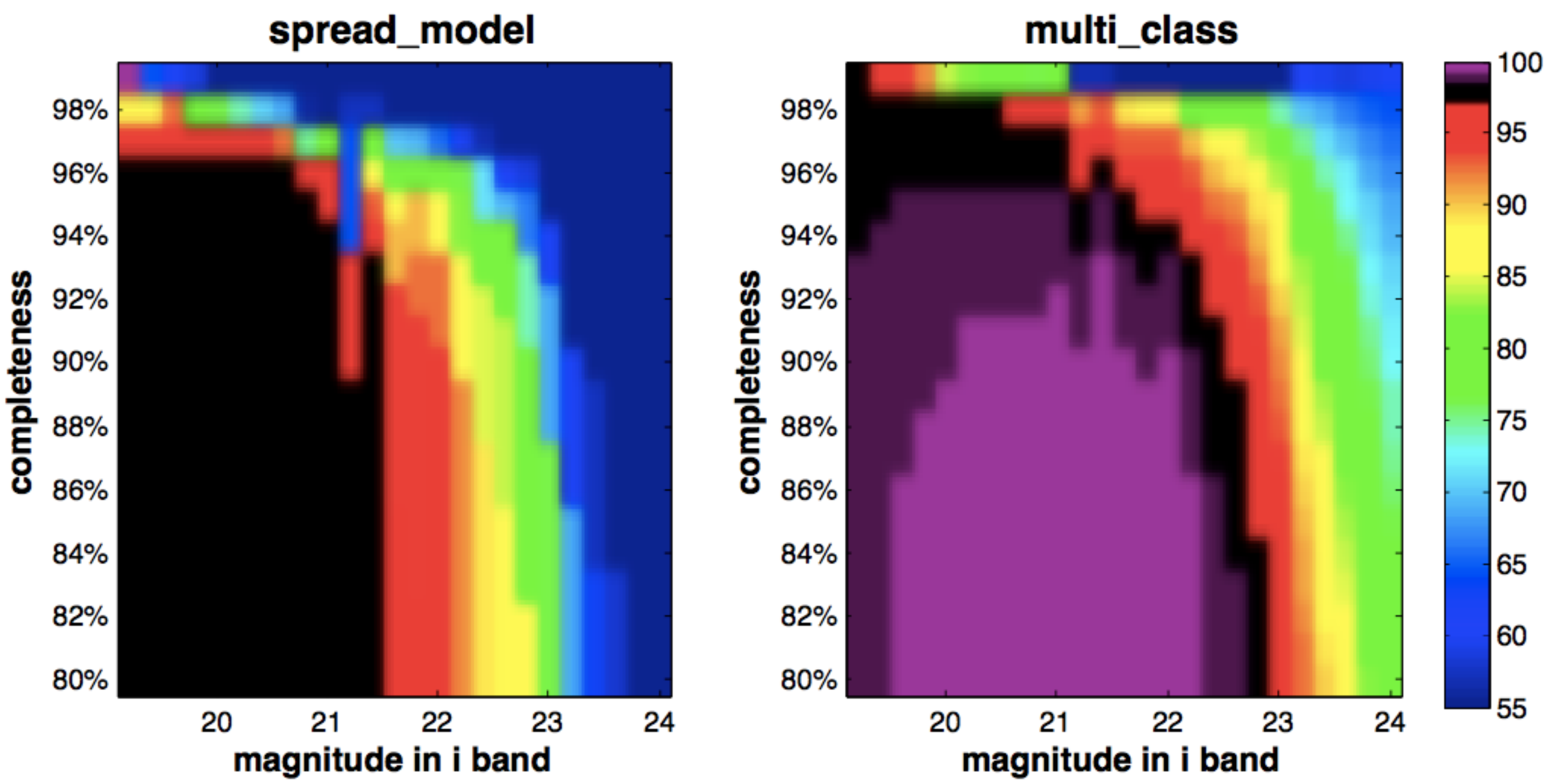}
\caption{Level of purity for a sample of stars $p^s$, for different magnitudes and values of the completeness. The $97\%$ science requirement (from WL, derived in section $3.4$) is shown in black. Higher purity levels are shown in purple and light purple. Our new estimator, multi\_class, allows us to widen the range of both magnitude and completeness where this requirement is achieved.}\label{fig:black_star}
\end{figure}

For LSS, our new classifier allows us to achieve requirements which cannot be fulfilled by spread\_model. Figure~\ref{fig:black_gal} shows that the $98.5\%$ limit on $p^g$ (derived in section $3.3.2$ and shown in purple on the figure) cannot be reached by spread\_model, whereas multi\_class allows us to reach it up to magnitudes of $22.9$ (at the required $88.9\%$ completeness level, derived in section $3.2.2$). 

\begin{figure}
\includegraphics[width=8.5cm]{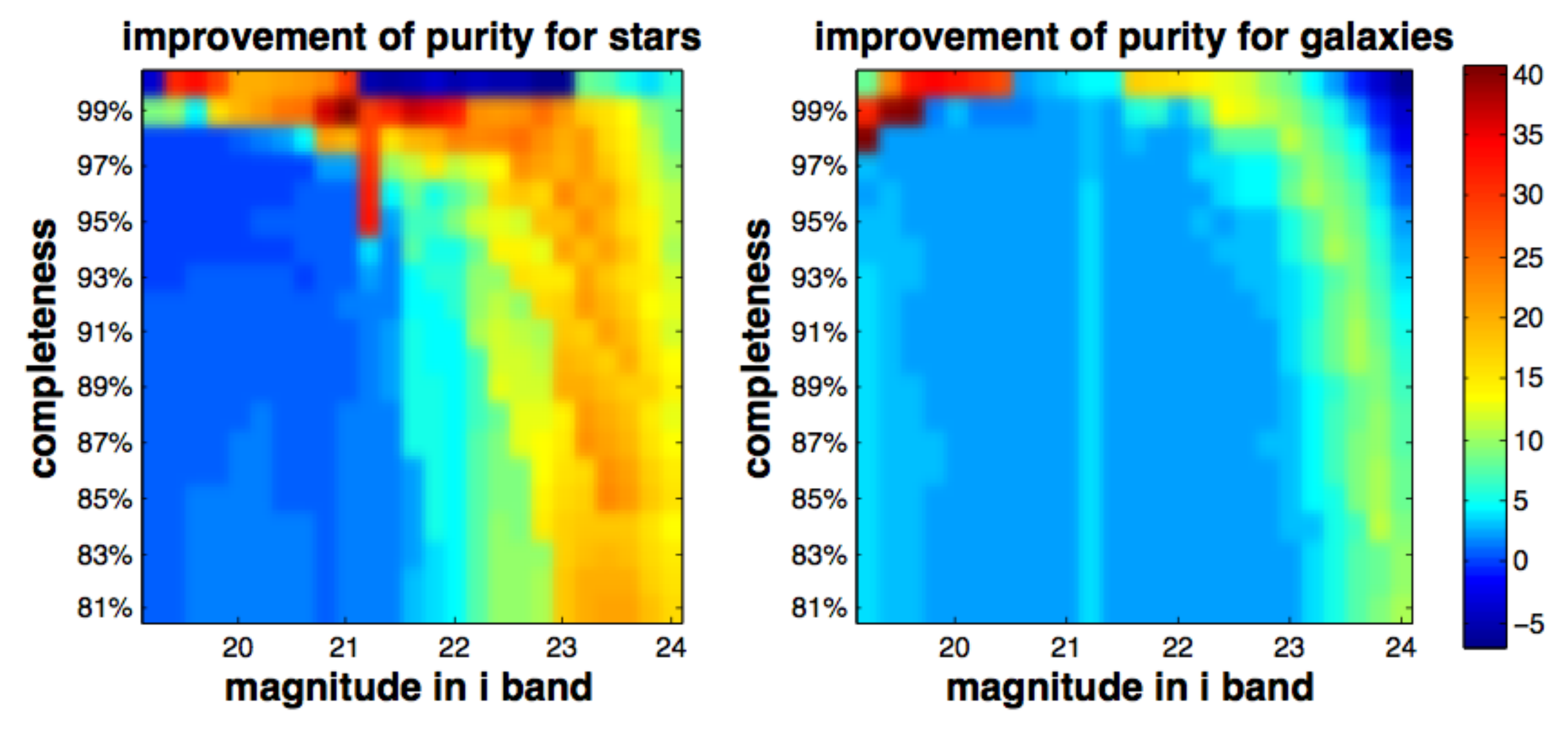}
\caption{Difference of the purity level achieved by multi\_class and spread\_model, $p_{multi\_class}-p_{spread\_model}$ for stars (left) and galaxies. At faint magnitudes (ranging from $23$ to $24$), multi\_class allows us to increase the level of $p^s$ achieved by spread\_model by up to $20\%$, and $p^g$ by up to $12\%$.}
\label{fig:impr}
\end{figure}

For WL, multi\_class allows us to increase the magnitude limit below which the science requirements are achieved. Figure~\ref{fig:Emmanuel} shows that this magnitude limit increases from $21.5$ to $23.4$ for the requirement on the stars purity $p^s$, and from $22.0$ to $22.9$ for the requirement on the galaxy purity $p^g$. Figure~\ref{fig:black_gal} and figure~\ref{fig:black_star} generalise this to a broad range of completenesses. 
In figure~\ref{fig:impr}, we consider the improvement in the purity of a sample of stars and a sample of galaxies, as a function of magnitude, for a large range of completenesses.
At faint magnitudes - typically fainter than 23 - multi\_class improves the purity achieved by spread\_model by up to $12\%$ for galaxies and by up to $20\%$ for stars. 


\section[]{Conclusions}

We showed that star/galaxy misclassification contributes to both the statistical and systematic error on the measurement of cosmological parameters. In particular, it affects the measurement of the DE equation of state parameters, $w_o$ and $w_a$, which future large photometric surveys such as DES aim to measure accurately. In the case of WL and LSS measurements, we translated the DETF FoM requirements on the statistical and systematic errors and the constraints from PSF calibration into the corresponding science requirements on the quality of star/galaxy separation. We formulated these requirements using two parameters: the purity and completeness of classified samples of stars and galaxy.

In order to meet these new requirements, we built an efficient method for star/galaxy classification, called multi\_class, which combines a PCA  with a learning algorithm. Our multi-parameter approach allows us to make use of the huge amount of information provided by {\it SExtractor}. In particular, the use of PCA allows us to better understand the correlations in the data, and to implement this physical knowledge in the classifier.

In ground-based surveys such as DES, the image quality is not constant with sky position and therefore any purely morphometric method gives limited performance, especially at faint magnitudes.  The flexibility of using an ANN allows us to consider the morphometry as one input parameters among many others and to integrate the performance of other classifiers to our new tool. Our new classifier, multi\_class, significantly improves the performance of the morphometric classifier implemented in {\it SExtractor} (spread\_model), which cannot achieve the LSS science requirements on star/galaxy separation. For both the LSS and WL probes, it allows us to widen the range of both magnitude and completeness where the derived science requirements are achieved. For magnitudes fainter than 23, multi\_class improves the purity achieved by spread\_model by up to $12\%$ for galaxies and by up to $20\%$ for stars. 

DES began survey operations in September, 2013, and will be running for five years. Therefore, we should be able to test the results shown in this paper on real data in the near future. The faint magnitudes reached by this new classifier constitute an important asset, which should allow to achieve the science requirements on star/galaxy separation in the next generation of wide-field photometric surveys. 

\section*{Acknowledgments}

MTS would like to thank Gary Bernstein, Benjamin Joachimi and Alan Heavens for very usefull comments and advice, Alexandre Refregier for a very useful discussion, and Ashley Ross, Adam Hawken, Manda Banerji, Alex Merson, Foteini Oikonomou, Boris Leistedt, Sreekumar Balan and Iftach Sadeh for their input to the project. MTS is grateful for the support from the University College London Perren and Impact studenships. FBA acknowledges the support of the Royal Society via a University Research Fellowship. OL acknowledges a Royal Society Wolfson Research Merit Award, a Leverhulme Senior Research Fellowship and an Advanced Grant from the European Research Council. BR acknowledges support from the European Research Council in the form of a Starting Grant with number 240672. We acknowledge UK's STFC for supporting DES optics and science. 

Funding for the DES Projects has been provided by the U.S. Department of Energy, the U.S. National Science Foundation, the Ministry of Science and Education of Spain, the Science and Technology Facilities Council of the United Kingdom, the Higher Education Funding Council for England, the National Center for Supercomputing Applications at the University of Illinois at Urbana-Champaign, the Kavli Institute of Cosmological Physics at the University of Chicago, Financiadora de Estudos e Projetos, Fundação Carlos Chagas Filho de Amparo à Pesquisa do Estado do Rio de Janeiro, Conselho Nacional de Desenvolvimento Científico e Tecnológico and the Ministério da Ciência e Tecnologia, the Deutsche Forschungsgemeinschaft and the Collaborating Institutions in the Dark Energy Survey.

The Collaborating Institutions are Argonne National Laboratories, the University of California at Santa Cruz, the University of Cambridge, Centro de Investigaciones Energeticas, Medioambientales y Tecnologicas-Madrid, the University of Chicago, University College London, DES-Brazil, Fermilab, the University of Edinburgh, the University of Illinois at Urbana-Champaign, the Institut de Ciencies de l'Espai (IEEC/CSIC), the Institut de Fisica d'Altes Energies, the Lawrence Berkeley National Laboratory, the Ludwig-Maximilians Universität and the associated Excellence Cluster Universe, the University of Michigan, the National Optical Astronomy Observatory, the University of Nottingham, the Ohio State University, the University of Pennsylvania, the University of Portsmouth, SLAC, Stanford University, the University of Sussex, Texas A\&M University, and the Institute of Astronomy at ETH-Zurich.

\appendix
\section[]{Statistical errors on the cosmological parameters \{$\Omega_m$, $H$, $\sigma_8$, $\Omega_b$, $n_s$, $b_g$\} from WL and LSS measurements}
\begin{figure*}
\includegraphics[width=11cm, height=7cm]{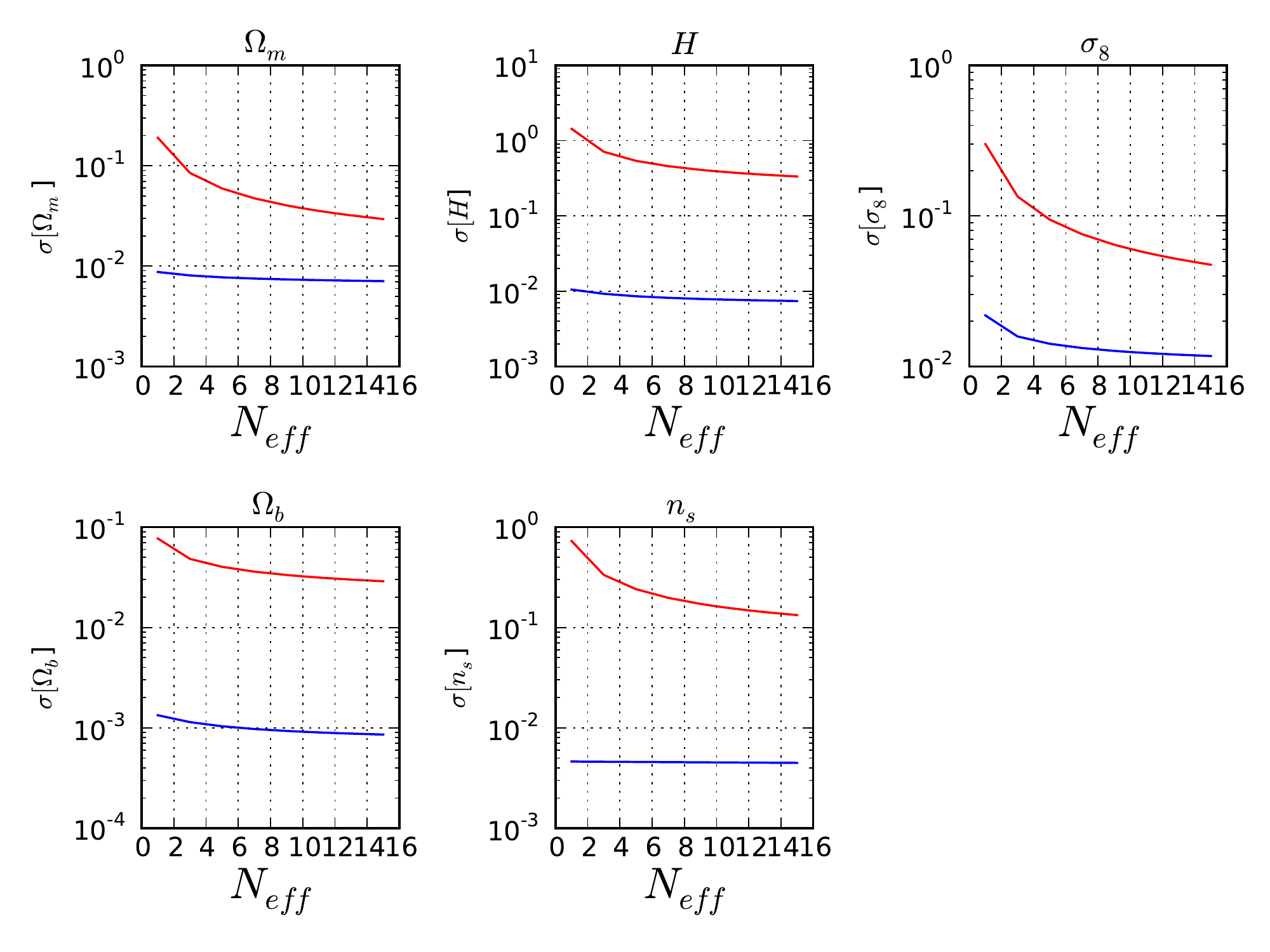}
\caption{Marginalised statistical errors on \{$\Omega_m$, $H$, $\sigma_8$, $\Omega_b$, $n_s$, $b_g$\} from the WL probe.}
\label{fig:neff_sigma_WL_others}
\end{figure*}

\begin{figure*}
\includegraphics[width=11cm,height=7cm]{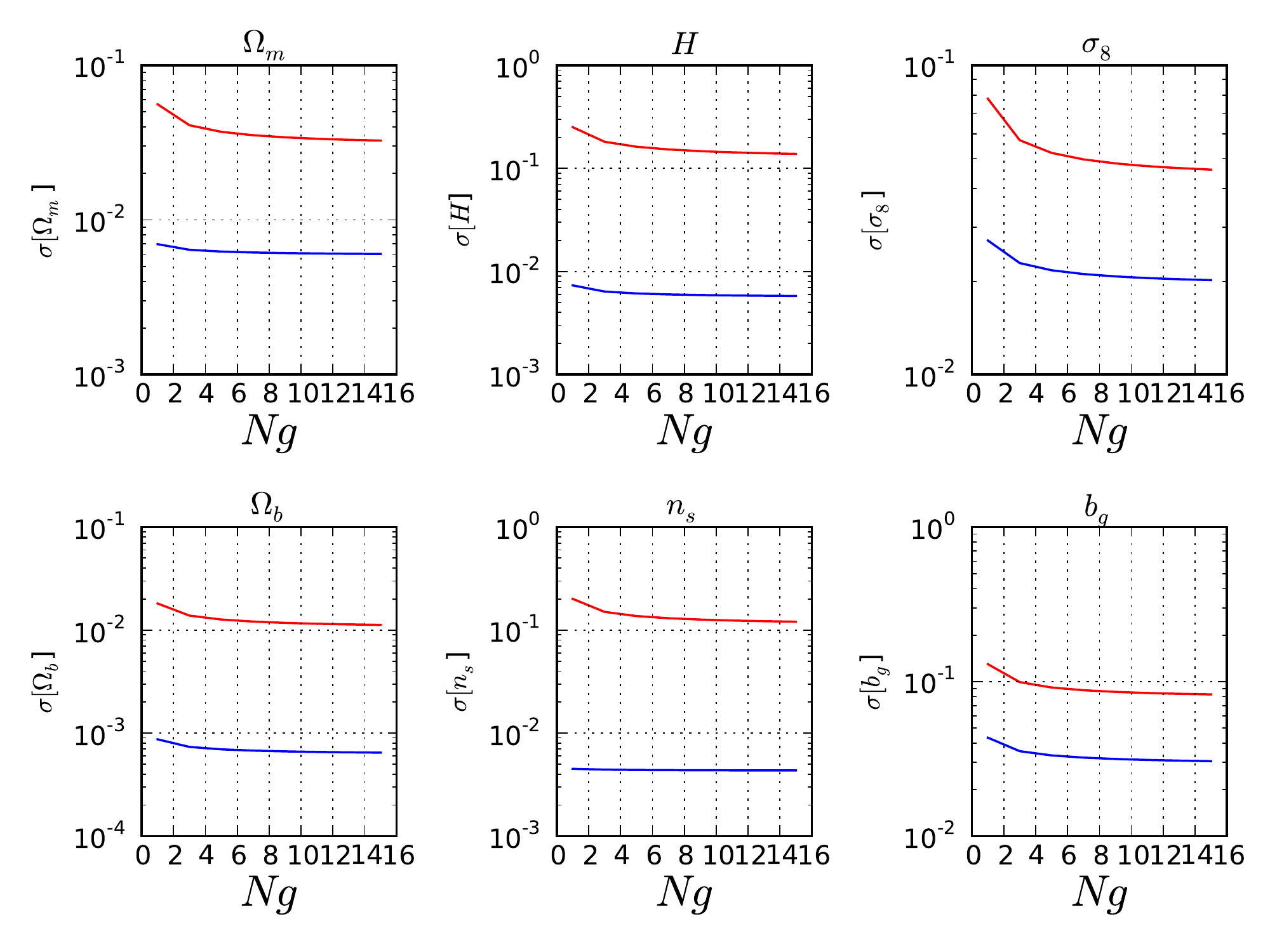}
\caption{Marginalised statistical errors on \{$\Omega_m$, $H$, $\sigma_8$, $\Omega_b$, $n_s$, $b_g$\} from the LSS probe.}
\label{fig:neff_sigma_LSS_others}
\end{figure*}

In the following, we show the marginalised statistical errors on \{$\Omega_m$, $H$, $\sigma_8$, $\Omega_b$, $n_s$, $b_g$\} from the WL probe (Figure~\ref{fig:neff_sigma_WL_others}) and the LSS probe (Figure~\ref{fig:neff_sigma_LSS_others}), for different values of the density of galaxies with reliable shape measurement $N_{eff}$ and of the density of detected galaxies $N_G$ respectively. The errors are marginalised and computed  using the assumptions and setup described in section $3.2.1$, with $l\in [1,1024]$ in the WL case and with $l\in [10, 400]$ in the LSS case. The red curve shows the errors computed with a non-informative prior whereas the blue curve is obtained assuming a Planck prior.


\bsp
\label{lastpage}

\end{document}